\documentclass[useAMS,usegraphicx,usenatbib]{mn2e} 
\usepackage{amsmath,graphicx}
\usepackage{ulem}
\usepackage{color}
\usepackage{graphicx}
\usepackage{times} 
\usepackage{amssymb}
\usepackage{amsmath}
\usepackage{lscape}
\usepackage{url}
\usepackage{multirow}
\newcommand{\kpc}{\rm\thinspace kpc}

\newcommand{\km}{\rm\thinspace km}

\newcommand{\cm}{\rm\thinspace cm}

\newcommand{\cmcups}{\hbox{$\cm^{3}\s^{-1}\,$}}
\newcommand{\yr}{\rm\thinspace yr}
\newcommand{\Gyr}{\rm\thinspace Gyr}

\newcommand{\s}{\rm\thinspace s}

\newcommand{\GHz}{\rm\thinspace GHz}

\newcommand{\Msun}{\hbox{$\rm\thinspace M_{\odot}$}}

\newcommand{\Msunpyr}{\hbox{$\Msun\yr^{-1}$}}
\newcommand{\keV}{\rm\thinspace keV}

\newcommand{\erg}{\rm\thinspace erg}

\newcommand{\ergpcmsqps}{\hbox{$\erg\cm^{-2}\s^{-1}$}}

\newcommand{\ergps}{\hbox{$\erg\s^{-1}\,$}}

\newcommand{\kmps}{\hbox{$\km\s^{-1}\,$}}

\newcommand{\asec}{\rm\thinspace arcsec}

\newcommand{\emm}{\hbox{$\cm^{-5}\,$}}
\newcommand{\empasecsq}{\hbox{$\emm\asec^{-2}\,$}}

\newcommand{\pcmsq}{\hbox{$\cm^{-2}\,$}}
\newcommand{\pcmcu}{\hbox{$\cm^{-3}\,$}}

\def\h0{\mbox{{\rm H}$^0$}}
\def\he0{\mbox{{\rm He}$^0$}}
\newif\ifAMStwofonts
\AMStwofontstrue
\newcommand{\dirlatexcommon}{/home/bcanning/Documents/latex_common/} 

\begin{document}

\title[AGN feedback in Abell 3581] {A multi-wavelength view of cooling vs. AGN heating in the X-ray luminous cool-core of Abell 3581\thanks{Based on observations obtained at the Southern Astrophysical Research (SOAR) telescope, which is a joint project of the Minist\'{e}rio da Ci\^{e}ncia, Tecnologia, e Inova\c{c}\~{a}o (MCTI) da Rep\'{u}blica Federativa do Brasil, the U.S. National Optical Astronomy Observatory (NOAO), the University of North Carolina at Chapel Hill (UNC), and Michigan State University (MSU).}} \author[R.E.A.~Canning] {\parbox[]{6.in}
  {R.E.A.~Canning$^{1,2}$\thanks{E-mail:
      rcanning@stanford.edu}, M.~Sun$^{3}$, J.S.~Sanders$^{4}$, T.E.~Clarke$^{5}$, A.C.~Fabian$^{6}$, S.~Giacintucci$^{7,8}$, D.V.~Lal$^{9}$, N.~Werner$^{1,2}$, S.W.~Allen$^{1,2,10}$,
      M.~Donahue$^{11}$, A.C.~Edge$^{12}$, R.M.~Johnstone$^{6}$, P.E.J.~Nulsen$^{13}$, P.~Salom\'{e}$^{14}$, C.L.~Sarazin$^{15}$\\ } \\
  \footnotesize
  $^{1}$Kavli Institute for Particle Astrophysics and Cosmology (KIPAC), Stanford University, 452 Lomita Mall, Stanford, CA 94305-4085, USA\\
  $^{2}$Department of Physics, Stanford University, 452 Lomita Mall, Stanford, CA 94305-4085, USA\\
  $^{3}$Eureka Scientific Inc., 2452 Delmer Street Suite 100, Oakland, CA 94602-3017, USA\\
  $^{4}$Max-Planck-Institut f\"{u}r extraterrestrische Physik (MPE), Giessenbachstrasse, 85748 Garching, Germany\\
  $^{5}$Naval Research Laboratory, 4555 Overlook Avenue SW, Code 7213, Washington, DC 20375, USA\\
  $^{6}$Institute of Astronomy, Madingley Road, Cambridge, CB3 0HA\\
  $^{7}$Department of Astronomy, University of Maryland, College Park, MD 20742-2421, USA\\
  $^{8}$Joint Space-Science Institute, University of Maryland, College Park, MD, 20742-2421, USA\\
  $^{9}$National Centre for Radio Astrophysics (NCRA-TIFR), Pune University Campus, Ganeshkhind P.O., Pune 411 007, India\\
  $^{10}$SLAC National Accelerator Laboratory, 2575 Sand Hill Road, Menlo Park, CA 94025, USA\\
  $^{11}$Michigan State University, 567 Wilson Rd. Rm 3275, East Lansing, MI 48824, USA\\  
  $^{12}$Institute of Computational Cosmology, University of Durham, Durham, DH1 3LE\\
  $^{13}$Harvard-Smithsonian Center for Astrophysics, 60 Garden Street, Cambridge, MA 02138, USA\\
  $^{14}$LERMA \& UMR8112 du CNRS, Observatoire de Paris, 61 Av. de l'Observatoire, F-75014 Raris, France\\
  $^{15}$Department of Astronomy, University of Virginia, 530 McCormick Road, Charlottesville, VA 22904, USA\\}
\maketitle

\begin{abstract}
We report the results of a multi-wavelength study of the nearby galaxy group, Abell 3581 ($z=0.0218$). This system hosts the most luminous cool core of any nearby group and exhibits active radio mode feedback from the super-massive black hole in its brightest group galaxy, IC 4374. The brightest galaxy has suffered multiple active galactic nucleus outbursts, blowing bubbles into the surrounding hot gas, which have resulted in the uplift of cool ionised gas into the surrounding hot intragroup medium. High velocities, indicative of an outflow, are observed close to the nucleus and coincident with the radio jet. Thin dusty filaments accompany the uplifted, ionised gas. No extended star formation is observed, however, a young cluster is detected just north of the nucleus. The direction of rise of the bubbles has changed between outbursts. This directional change is likely due to sloshing motions of the intragroup medium. These sloshing motions also appear to be actively stripping the X-ray cool core, as indicated by a spiraling cold front of high metallicity, low temperature, low entropy gas.
\end{abstract}


\section{Introduction}

\begin{figure*}
\centering
\includegraphics[width=0.495\textwidth]{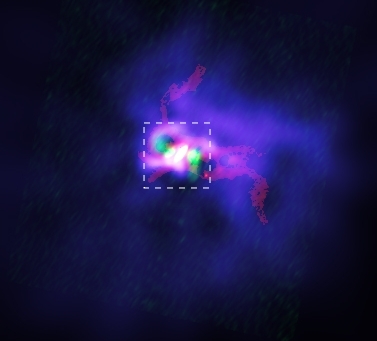}
\hspace{0.1cm}
\includegraphics[width=0.448\textwidth]{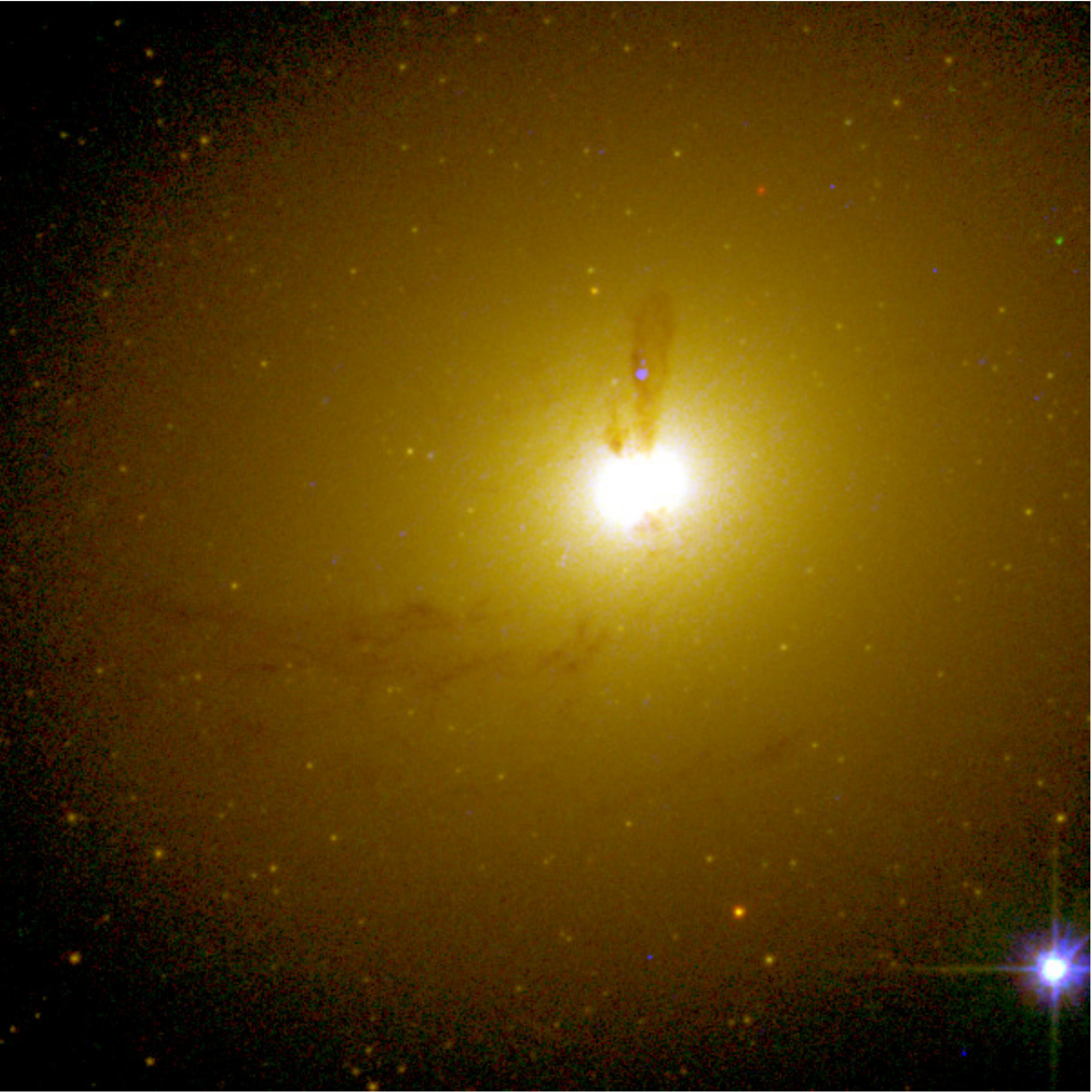}
\caption{{\bf Left:} Three colour image of Abell 3581 formed from an adaptively smoothed
\textit{Chandra} X-ray image with the average at each radius removed (blue), archival VLA
1.4~GHz radio emission (green) and SOAR narrowband H$\alpha$+[N {\sc ii}] image
(red). The radio bubbles push aside the hot X-ray gas forming the inner two
cavities. Another `ghost' X-ray cavity is observed to the south-west devoid of
radio emission. Filaments of ionised gas encase the young bubbles and are
entrained beneath the older cavity. {\bf Right:} Colour composite of the central group galaxy, IC 4374, in the three HST filters F300X (blue), F555W (green) and F814W (red). The region shown to the right is highlighted on the left hand figure by the white rectangle which has dimensions 24.2 arcsec (10.5~kpc) across. Dust filaments are observed aligned with the H$\alpha$+[N {\sc ii}] and a young star cluster is observed in the dust to the north of the nucleus. \label{3_colour}} 
\end{figure*}

The last decade has seen significant progress into the cooling flow phenomenon in `cool core' clusters of galaxies. These clusters are so named as they exhibit sharply peaked X-ray surface brightness profiles resulting from their cool, high density X-ray cores. The cooling time of the gas in these cores is significantly shorter than the Hubble time. 

If no sources of heat are present, the observed high central X-ray luminosities would imply mass cooling rates as high as hundreds or thousands of solar masses a year (e.g. \citealt{fabian1994} and references therein). However, with the advent of the XMM-\textit{Newton} space telescope, high resolution spectra exposed a lack of strong X-ray emission below 1 \keV\ in cool cores clusters. The implication of these observations was a reduction in the inferred net X-ray cooling rates and an effective quenching of any resulting star formation (for reviews see \citealt{peterson2006} and \citealt{mcnamara2007}). The most probable source of heating to quench the cooling of the intracluster medium (ICM) is provided by outbursts from the super-massive black hole (SMBH) lurking at the centre of the central, dominant, brightest cluster galaxies (BCGs) (for a review of the observational evidence for BH feedback see \citealt{fabian2012}).

The prevalence of H$\alpha$ emission in BCGs has long been known (e.g. \citealt{lynds1970, cowie1983, hu1983, johnstone1987, heckman1989, crawford1999, donahue2000,fabian2001, conselice2001}) and has been shown to have diverse and dramatic morphologies. These ionised gas nebulosities surround about a third of all BCGs but are found preferentially in cool core clusters. The luminosity of the H$\alpha$ bright nebulosities ranges from $10^{39}-10^{43}$ \ergps\ and can extend over 70 \kpc\ from the core of the BCG \citep{crawford1992, crawford1999}. The emission is well correlated with both the X-ray luminosity and with emission lines from molecular gas. Developments in optical astronomy, particularly with the progression of integral field spectroscopy (IFS), have allowed the spatio-kinematic properties of the gaseous nebulae surrounding BCGs to be better probed. For example, the spatial, kinematical and excitation properties of the ionised and molecular gas has now been mapped in small samples of BCGs (e.g. \citealt{wilman2006, hatch2007,wilman2009, edwards2009, farage2010, oonk2010, canning2011b, farage2012}).

In addition to ionised gas nebulosities, many cool core BCGs contain significant masses of cold molecular gas and dust, both in the centres of the galaxies and farther out. This is often coincident with the extended ionised gas filaments in the galaxy outskirts (e.g. \citealt{jaffe1997, donahue2000, edge2001, salome2003, hatch2005, salome2006,johnstone2007, oonk2010, edge2010a, mittal2011, werner2012}). Answering the questions of what the origin of these cool ($\sim10^{3}-10^{4}$~K) and cold ($<10^{3}$~K) gas phases are and how the complex heating and cooling balance in these systems is maintained, requires observations of multiphase material; from synchrotron emitting relativistic plasmas to cold molecular gas and dust. 

\begin{table*}
 \centering
 \caption{Summary of observations. }
 \begin{tabular}{c|c|c|c|c|c}
  Telescope & Filter & Exposure & Date & ID & PI \\
\hline \hline
  SOAR/SOI & CTIO 6693/76 & 3$\times$600~s & 2008 July 8 & & Sun/Donahue \\
   & CTIO 6520/76 & 3$\times$360~s & 2008 July 8 & & \\
  HST/WFC3/UVIS & F300X & 5427~s & 2011 February 2nd & 12373 & Sun \\
   & F555W & 1335~s & 2011 February 2nd & & \\
   & F814W & 1044~s & 2011 February 2nd & & \\
  VLT/VIMOS & HRO grism, 0.67'' fibers & 4500~s & 2009 March 30th & 082.B-0093(A) & Fabian \\
\hline
  JVLA & L band, BnA configuration & 4~hr & 2012 September 8th & SB0410 & Sun \\
  GMRT & 610~MHz & 2.4~hr & 2010 May & 18-009 & Sun \\
       & 235~MHz & 2.4~hr & 2010 May & & \\ 
\hline
  Chandra/ACIS-S &  & 85~ks & 2011 January 3rd & 12884 & Sun \\
\hline
 \end{tabular}
 \label{obs_table}
\end{table*}

Groups and poor clusters have shallower potential wells compared with rich galaxy clusters. An important implication of this is that the mechanical output from an active galactic nucleus (AGN) outburst is of the same order as the group binding energy. The effects of AGN feedback in groups may well be different to that observed on the cluster scale. The majority of galaxies in the Universe are situated in group environments. Therefore, investigating the nature of AGN feedback in lower mass systems is paramount to our understanding of galaxy evolution. 

The nearby galaxy group, Abell 3581 (Fig. 1, $z=0.0218$, distance $d=93.6$~Mpc), presents an excellent case study for the investigation of the thermal
regulation of groups of galaxies. It has a system temperature of $\sim$ 1.7 keV \citep{sun2009a} and hosts the most luminous X-ray cool core among galaxy groups with $z\leq$0.05. The central AGN, PKS 1404-267, has the highest radio power of any nearby group with a large cool core (e.g. \citealt{sun2009b}). This cool core co-exists with one of the most spectacular and luminous H$\alpha$ nebulae observed in galaxy groups, which surrounds the brightest group galaxy (BGG) IC 4374 \citep{johnstone1987,
danziger1988}. CO$(2-1)$ emission with an intensity of 0.62$\pm$0.15 K\kmps\ ($I_{CO}=\int T\Delta V$), is also detected in the central galaxy. The detection of CO$(2-1)$ was made with IRAM-30m observations of the nucleus of Abell 3581 (the observations were centered on RA$14^{h}07^{m}29^{s}.76$, Dec. $-27^{\circ}$ 01' 05.9'', J2000) with an 11 arcsec full width half maximum (FWHM) beam width. The previous \textit{Chandra} observations reported in \cite{johnstone2005} have shown that two X-ray cavities exist in the centre of the system, where the AGN jets are currently displacing the hot interstellar medium (ISM). \cite{sanders2010} observed Fe {\sc xvii} emission lines in the core of Abell 3581, indicating the presence of low temperature, 0.5~\keV\ X-ray emitting gas. The authors find evidence that this X-ray coolest detected gas phase has a low volume filling fraction (3 per cent) so the cool blobs exist interspersed in the hotter surrounding medium. Such low volume filling factor cool X-ray gas has been seen in other objects (e.g. \citealt{sanders2007, sanders2009, sanders2010, werner2010, werner2012}). \cite{sanders2010} find no evidence for O {\sc vii} emission lines in Abell 3581 implying very little of the gas can be cooling radiatively below 0.5~\keV. Either non-radiative cooling is occurring or a heating source is needed which targets this coolest X-ray gas without overheating the surroundings. Thus, Abell 3581 is an ideal system in which to study AGN heating and the multi-phase gas in group cool cores.

In this paper we present deep multi-wavelength imaging and spectroscopy of the central galaxy in Abell 3581 and the surrounding ICM (details of the observations are given in Table. \ref{obs_table}). Throughout this paper we assume the standard $\Lambda$CDM cosmology where H$_{0}$ = 71 km~s$^{−1}$, $\Omega_{m}$ = 0.27 and $\Omega_{\Lambda}$ = 0.73. For this cosmology and at the redshift of IC 4374 ($z=0.0218$), an angular size of 1'' corresponds to a distance of 0.435 kpc.\\

\section{Optical data}

\subsection{SOAR}

Narrow-band optical imaging was performed with the 4.1~m Southern Astrophysical Research (SOAR)
telescope on July 8, 2008 (UT) using the SOAR Optical Imager
(SOI). The night was photometric with a seeing of around 1.0$''$.
The CTIO 6693/76 narrow-band filter was used for the H$\alpha$+[N {\sc ii}]
lines, and the CTIO 6520/76 filter for the continuum. Three 600
sec exposures were taken with the 6693/76 filter and three 360 sec
exposures with the 6520/76 filter. Each image was reduced using
the standard procedures in the IRAF MSCRED package and the
spectroscopic standard EG~274. The pixels were binned 2$\times$2, for
a scale of 0.154 arcsec per pixel. More details on the SOI data reduction
and the continuum subtraction can be found in \cite{sun2007}. The net (continuum subtracted) H$\alpha$ image has been published in \cite{sun2012}.

\subsection{HST}

Observations were made using the Wide Field Camera 3 (WFC3) UVIS channel on the Hubble Space Telescope (HST) on 2011 February 2 in the F300X,
F555W and F814W filters. The total exposure times in each filter were 5427, 1335
and 1044 s respectively. Throughput of the F555W filter at the redshifted
wavelengths of the strong emission lines, H$\alpha+$[N {\sc ii}]
$\sim$6700$\mathrm{\AA}$, is only of order a few percent. Standard procedures
were used to bias-subtract and flat-field the data frames and to drizzle the
science frames together, these procedures use the software DrizzlePac and are described in the HST WFC3 data
handbook.

\begin{figure*}
\centering
\includegraphics[width=0.7\textwidth]{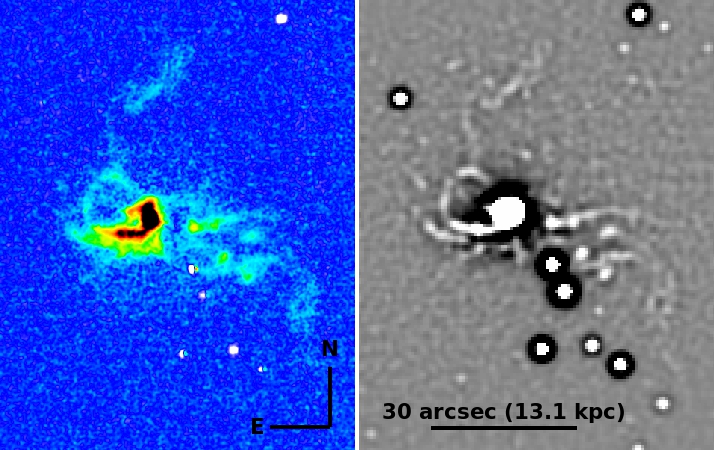}
\caption{{\bf Left:} Net narrow band H$\alpha$ emission image, taken with the SOAR telescope, of the BGG in Abell 3581
after removal of continuum emission with the CTIO 6520/76 filter. {\bf Right:}
The unsharp masked H$\alpha$ emission image. The initial image has been smoothed
by a 2D gaussian with $\sigma=1$~arcsec, subtracted from the original and binned
by 4 pixels (0.6'') in each direction.
\label{ha_image}}
\end{figure*}

\begin{figure*}
\centering
\includegraphics[width=0.9\textwidth]{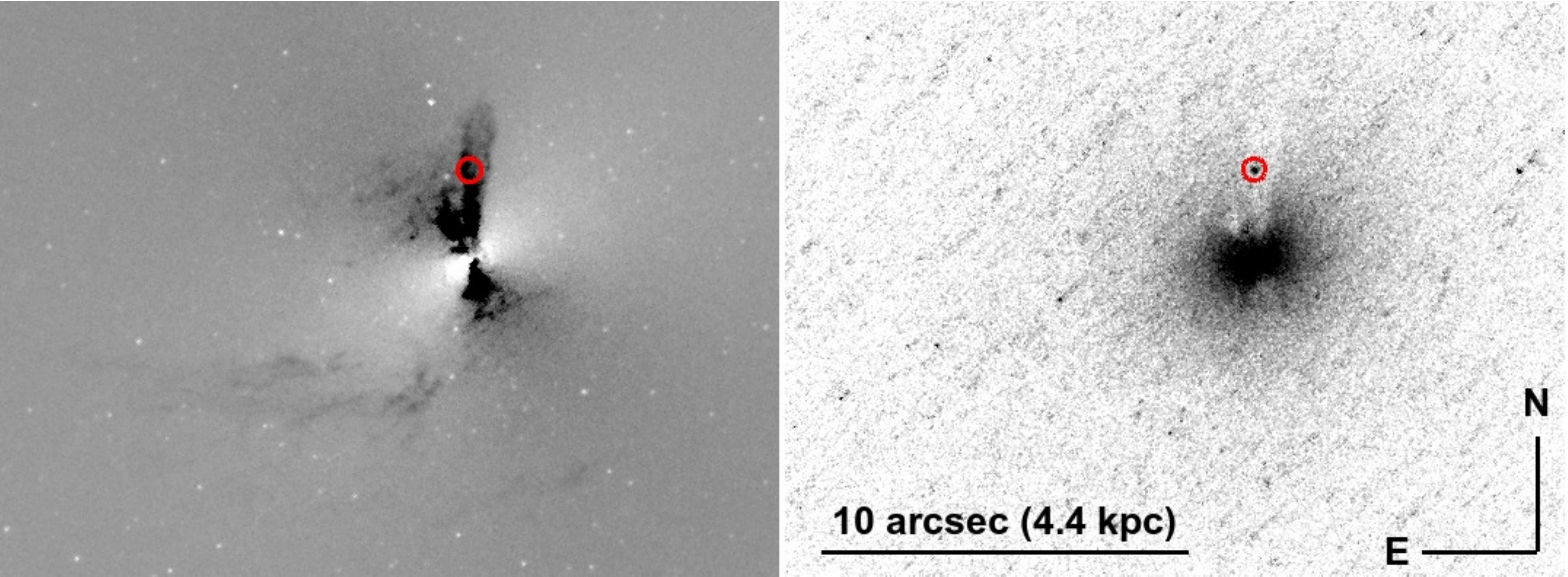}
\caption{{\bf Left:} HST F555W image with a smooth elliptical isophotal model subtracted. The dark
feature piercing north-south through the nucleus and turning towards the
south-east is absorption from a prominent dust lane. {\bf Right:} F300X image of
the same region. The red circle shows a young star cluster not seen in the
visible (F555W and F814W) images. The star cluster coincides with the northern
dust lane. \label{dust_and_stars}}
\end{figure*}

\begin{figure*}
\centering
\includegraphics[width=0.9\textwidth]{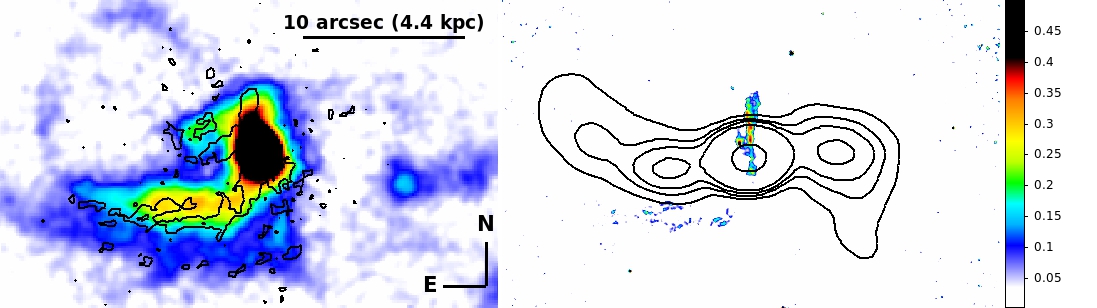}
\caption{{\bf Left:} The central region of the net H$\alpha$ image overlaid with contours of
dust absorption. The dust closely coincides with the inner H$\alpha$
filaments. {\bf Right:} The relative dust attenuation in the brightest cluster
galaxy in units of $E(B-V)$ magnitudes after correction for Galactic reddening.
Contours of L-band JVLA radio emission, which fill the inner X-ray cavities, are overlaid. The angular scale is the same as the left hand panel. \label{dust_and_halpha}}
\end{figure*}

The optical images are shown in Figs. \ref{ha_image}, \ref{dust_and_stars} and
\ref{dust_and_halpha}. No corrections for reddening have been made to the
images, although a relative intrinsic $E(B-V)$ image of the galaxy is shown in the
right-hand panel of Fig. \ref{dust_and_halpha}. This reddening map is formed by first correcting the broad band HST imaging for Galactic
reddening in the direction of Abell 3581. We take the Galactic reddening to be $E(B-V)=0.060\pm0.001$ using the reddening maps of \cite{schlegel1998} and assuming the
Galactic extinction law of \cite{cardelli1989}. We then converted the HST/WFC3/UVIS broad band images into AB
magnitude units and subtracted the F814W broad band HST image from the F555W
band. We then subtracted the value at a reference point, within the galaxy, but
where there was no evidence for dust lanes in the single F555W or F300X frames.
The point chosen as our reference is given by the coordinates RA
$14^{h}07^{m}29^{s}.768$, Dec. $-27^{\circ}$ 01' 05.20'' (J2000). This then
provides an $E(\mathrm{F555W}-\mathrm{F814W})$ relative intrinsic reddening map. 

To convert the map to $E(B-V)$ units requires the assumption of an attenuation
curve. We assume the \cite{cardelli1989} extinction law and calculate
A$_{\lambda}$/A$_{V}$ from the HST filters. We use the pivot wavelengths of the
F555W and F814W filters given in the HST WFC3 data handbook as the effective
wavelength of the filters. Due to the broad nature of the filters and the low
redshift of our object, no K-correction is made.

\subsection{Integral field spectroscopy}
\label{ifu}

Observations were made on 2009 March 30 using the VIsible MultiObject
Spectrograph (VIMOS) on the Very Large Telescope (VLT) in Cerro Paranal, Chile (see \citealt{lefevre2003}
and \citealt{zanichelli2005} for a description of the VIMOS IFU and a discussion
of data reduction techniques). We obtained High Resolution Orange (HRO,
$R=2650$) data using the larger 0.67'' fibers giving a field-of-view of 27''$\times$27'' with 1600 fibers. The wavelength coverage across the whole field-of-view in the HRO
grism is from $5250-7400~\mathrm{\AA}$ which cover the redshifted spectral lines
of [N {\sc i}]$\lambda$5199 to [S {\sc ii}]$\lambda$6731, including the strong spectral lines of [O {\sc i}]$\lambda$6300, H$\alpha$ and [N {\sc ii}]$\lambda$6548,6583. However, due to the   centring of the pseudo-slits in each quadrant the spectral coverage varies from quadrant-to-quadrant. The result is that the eastern side of our field-of-view has additional coverage of the redshifted spectral line of [O {\sc iii}]$\lambda$5007. The total
exposure time of these observations was 4500 seconds.  

The data were reduced by the {\sc vipgi}\footnote{VIMOS Interactive
Pipeline Graphical Interface (VIPGI), obtained from
http://cosmos.iasf-milano.inaf.it/pandora/.} pipeline (see
\citealt{scodeggio2005} for a description of {\sc vipgi}). The data cubes were
combined, transmission corrected, sky absorption and emission corrected and
corrected for Galactic extinction ($E(B-V)=0.060\pm0.001$) using a set of IDL
routines. The O$_{2}$ telluric absorption feature at
$\sim~6800-6900~\mathrm{\AA}$ falls over the [S {\sc ii}] doublet and as such
warranted special attention. We use observations of two standard stars taken
just before and after the science observations to correct for this feature. 

Due to the nature of the filamentary system we choose to bin the data based on
the surface brightness of [N {\sc ii}]$\lambda$6583 emission, the strongest line
observed, using the contour binning technique of \cite{sanders2006c} and binning
to a signal-to-noise of 10.  

We then fit the emission lines in IDL using {\sc mpfit} \citep{more1978,
markwardt2009}. The strong emission lines of [N {\sc ii}]$\lambda$6548,6583 and H$\alpha$ are fit simultaneously
across the field-of-view. The redshift and velocity dispersion are constrained to be
the same for these emission lines and the integrated flux of the [N {\sc ii}]
doublet is tied in the ratio 3:1, the scaling being dictated by the atomic
parameters \citep{osterbrock2006}. The [S {\sc ii}] and [O {\sc i}] lines are
fit separately (the [O {\sc i}] doublet flux is also tied by atomic parameters);
however, the morphology and velocity structure of all emission lines are similar.
We fit gaussians to the spectra with 1 and 2 velocity components and both with
and without a broad H$\alpha$ component. Using {\sc mpftest} in IDL we determine that
a one component velocity structure is sufficient to fit the majority of our data.
However, we require a broad H$\alpha$ component within the inner 2 square arcseconds
radius of the galaxy.

In the images from IFU data, only pixels where the emission lines were detected at a 4 $\sigma$ or greater level are shown. The instrumental spectral resolution is determined from a fit to strong sky emission lines and is found to be 98 \kmps\ at full width half maximum (FWHM).

\section{X-ray data}
\label{x-ray_data}

\textit{Chandra} X-ray data of Abell~3581, with an exposure time of 85~ks, were taken on 2011
January 3rd (ObsID 12884). The observations were conducted with the Advanced CCD Imaging Spectrometer (ACIS).
The data was reprocessed using the \textsc{ciao} \citep{fruscione2006} 
\textsc{acis\_process\_events} tool, applying the VFAINT processing 
to lower the non X-ray background. To look for contaminating flares, we 
examined the lightcurve of ACIS CCD 5 between 2.5 and 7 keV in 200s 
bins. A sigma clipping algorithm was used to search for flared periods, of 
which there were none. Standard blank-sky background datasets were used, 
re-projected to match the observation and with the exposure time adjusted 
to match the background-dominated 9 to 12 keV observational count rate. 
To make the spectral maps, we used the Contour Binning algorithm 
\citep{sanders2006c} to create regions with a signal to noise ratio of 20 
(around 400 counts). Spectra and background spectra were extracted from 
these bins, and response and ancillary response regions were created, 
weighting response regions by the number of counts in the 0.5 to 7 keV 
band. We fitted the spectra in Xspec between 0.5 and 7 keV, minimising 
the modified C statistic. We used the APEC thermal model to fit the 
spectrum \citep{smith2001}, with Galactic absorption (4.36$\times$10$^{20}$~\pcmsq) applied. In the fits the 
temperature, metallicity and normalisation were allowed to vary.

In order to examine the surface brightness edges in the images we fit and subtracted a double-$\beta$ model, using the {\it Sherpa} fitting package, to an azimuthally averaged, background subtracted surface brightness profile, centred on and extracted between a radius of 2.5-120 arcsec. The surface brightness profile is from the exposure-corrected image with point sources removed manually. The image is in the energy range 0.5-7~keV. Despite the asymmetry of the cluster, the model fits the data reasonably well. However, deviations from the fit are seen at $\sim$50 arcsec, $\sim$80 arcsec, and in the central 20 arcsec. These deviations correspond to surface brightness edges which will be examined further in Section \ref{xray_section}. Double-$\beta$ fits to individual sectors and empirical azimuthally averaged profiles are also fit and removed from the surface brightness images to highlight substructure.

For the temperature and density profiles across the `bubble' and `cold front' (see Section \ref{xray_section}) we extracted spectra from the regions of interest, and fitted the spectrum within Xspec (version 12.8.0a\footnote{We used the AtomDB 2.0 release which includes changes to the iron L-shell data affecting kT $<$ 2 keV plasma which has the effect of increasing the temperature (10-20 per cent) while decreasing the abundances ($\sim$20 per cent) when compared with AtomDB 1.3.1.}) with an
absorbed {\sc mekal} model. The regions for the spectral analysis were chosen such
that there were at least 3000 counts in each bin. Only data in the 0.5-7.0~keV range were used. C statistics were used for the fit
and spectra were deprojected using the routine developed in \cite{sanders2007}
and \cite{russell2008}.  
The Galactic absorption in
the direction of Abell 3581 was taken as 4.36$\times$10$^{20}$~\pcmsq, using the value of
\cite{kalberla2005} and was frozen in the fits.  The solar photospheric abundance table by
\cite{anders1989} was used in the spectral fits.  Error bars on the temperature, density and pressure plots are shown at the 90 per cent level.

\section{Radio data}

\subsection{JVLA}

High frequency radio observations of Abell 3581 were obtained with the
NRAO\footnote{The National Radio Astronomy Observatory is a facility
of the National Science Foundation operated under cooperative
agreement by Associated Universities, Inc.} JVLA \citep{perley2011} at L
band (1-2 GHz). Due to the low declination of the target, the
observations were undertaken in the hybrid BnA configuration on 2012
Sept.\ 08 (project ID: SB0410). The extended north arm of the hybrid
configuration allows for a more circular beam for the target. The
observations were a total of 4 hours and were taken with two 512 MHz
wide basebands centered at 1314 and 1686 MHz. We configurated the
WIDAR correlator so that each baseband had 8 subbands with 64 channels
per subband for a frequency resolution of 1 MHz. We used 3C286 as our
absolute flux and bandpass calibrator and J1351-1449 as our phase
calibrator.

The data were flagged and calibrated using NRAO's Common Astronomy
Software Applications (CASA\footnote{http://casa.nrao.edu}) data
reductions package. We Hanning smoothed the data to reduce the effects
of Gibbs ringing around strong radio frequency interference (RFI)
which is common at L band. We then flagged the data in the time and
frequency domains to remove the majority of the significant
RFI. Additional time-domain flagging on the bright calibrators was
done using the average calibrator amplitudes.  We used all calibrators
to calculate the parallel hand group delays and applied the results to
all sources. We then followed standard techniques for bandpass
calibration, taking into acount the spectral index of the bandpass
calibrator, as well as gain calibration. Following calibration, we
split off the target field and undertook additional flagging before
imaging the data with MFS algorithm in CASA which allows us to deal
with the wideband data. We applied four rounds of phase only
self-calibration on the target followed by a round of amplitude and
phase self-calibration. The final image of Abell 3581 has a noise of
$\sigma = 22 \mu$Jy/bm with a resolution of 2.56 $\times$ 2.07~\arcsec\
at a position angle of $-84.95$ degrees.
 
\subsection{GMRT}

\begin{figure*}
\centering
\includegraphics[width=0.8\textwidth]{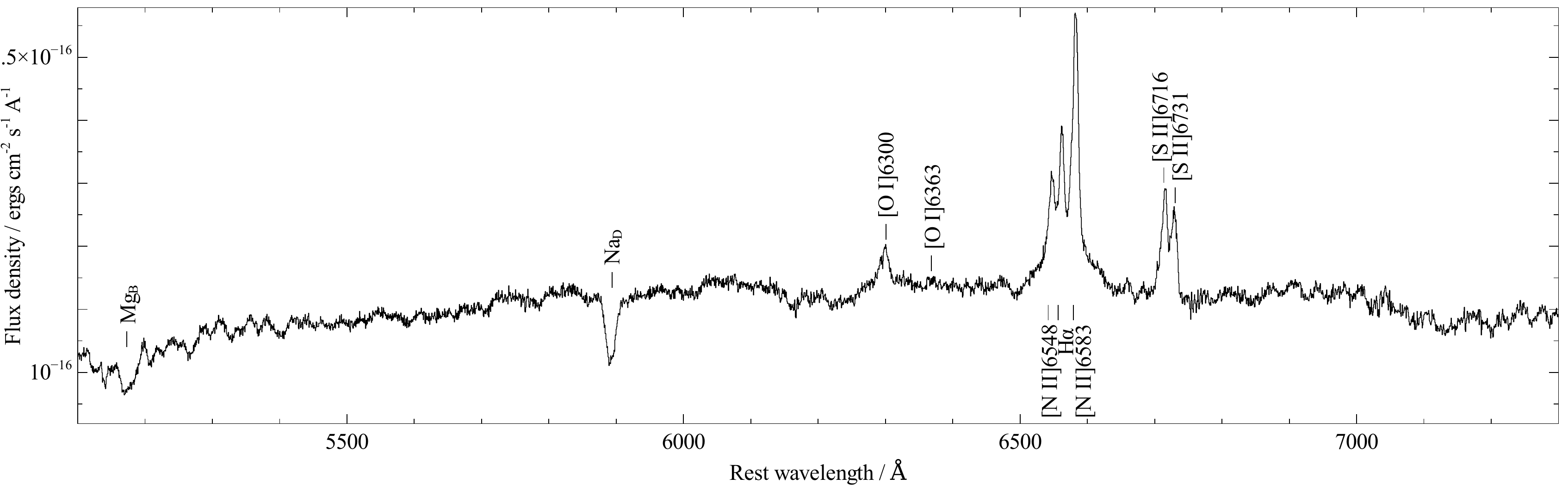}
\caption{The optical spectrum of the BGG nucleus in Abell 3581. \label{spectrum}}
\end{figure*}

Abell 3581 was observed with the GMRT at 235 MHz and 610 MHz (project 18-009) in May 2010, for approximately 3 hours, including calibration overheads. The observations were made using the dual-frequency mode. The 610 MHz data were collected using the upper and lower side bands (USB and LSB) with an observing bandwidth of 16 MHz each. Only the USB was used at 235 MHz, with a bandwidth of 8 MHz. We chose the default spectral-line observing mode, with 128 channels for each band at 610 MHz (64 channels at 240 MHz) and a spectral resolution of 125 kHz/channel. The datasets were calibrated and reduced using the NRAO Astronomical Image Processing System package (AIPS) as described in \cite{giacintucci2011}. Self-calibration was applied to the data to reduce residual phase variations and improve the quality of the final images. Due to the large field of view of the GMRT at low frequencies, we used the wide-field imaging technique at each step of the phase self-calibration process, to account for the non-planar nature of the sky. The final images were produced using the multi-scale CLEAN implemented in the AIPS task IMAGR, which results in better imaging of extended sources compared to the traditional CLEAN (e.g., \citealt{wakker1988,cornwell1993}). The full resolution of our final images is $8^{\prime\prime}\times4^{\prime\prime}$ at 610 MHz and $16^{\prime\prime}\times11^{\prime\prime}$ at 235 MHz. The rms noise level (1$\sigma$) in these images is 0.1 mJy beam$^{-1}$ and 0.9 mJy beam$^{-1}$, respectively.  Residual amplitude errors are within 8 per cent at 240 MHz and 5 per cent at 610 MHz.

\section{Results}

\subsection{Optical}

\subsubsection{Morphology}

\begin{figure*}
\centering
\includegraphics[width=0.4\textwidth]{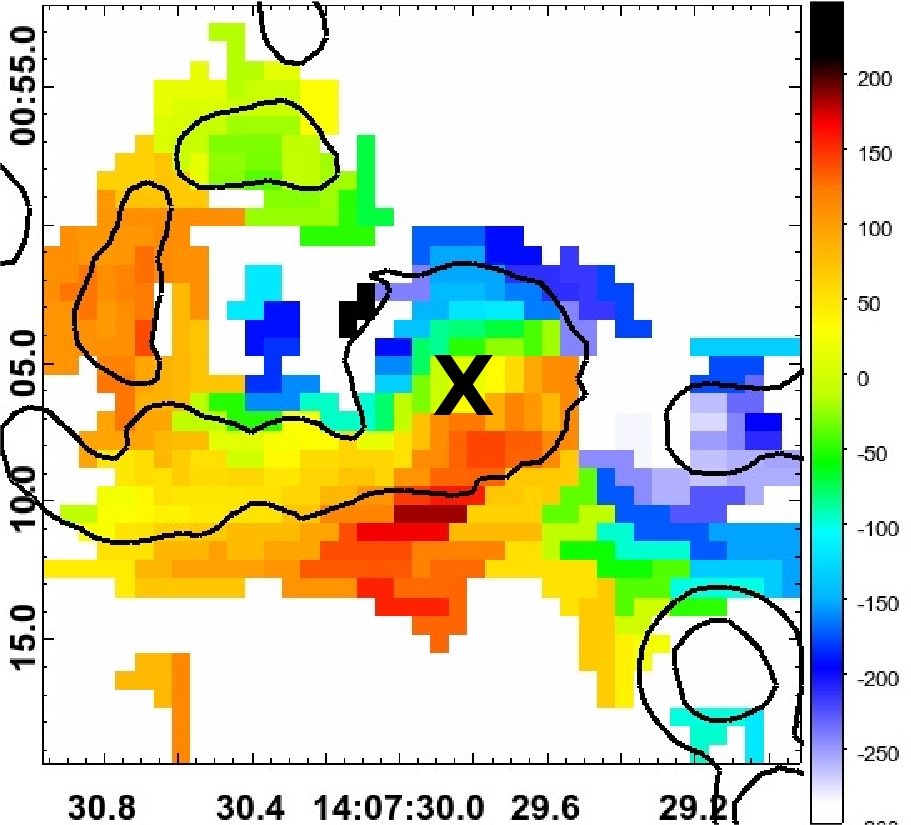}
\includegraphics[width=0.4\textwidth]{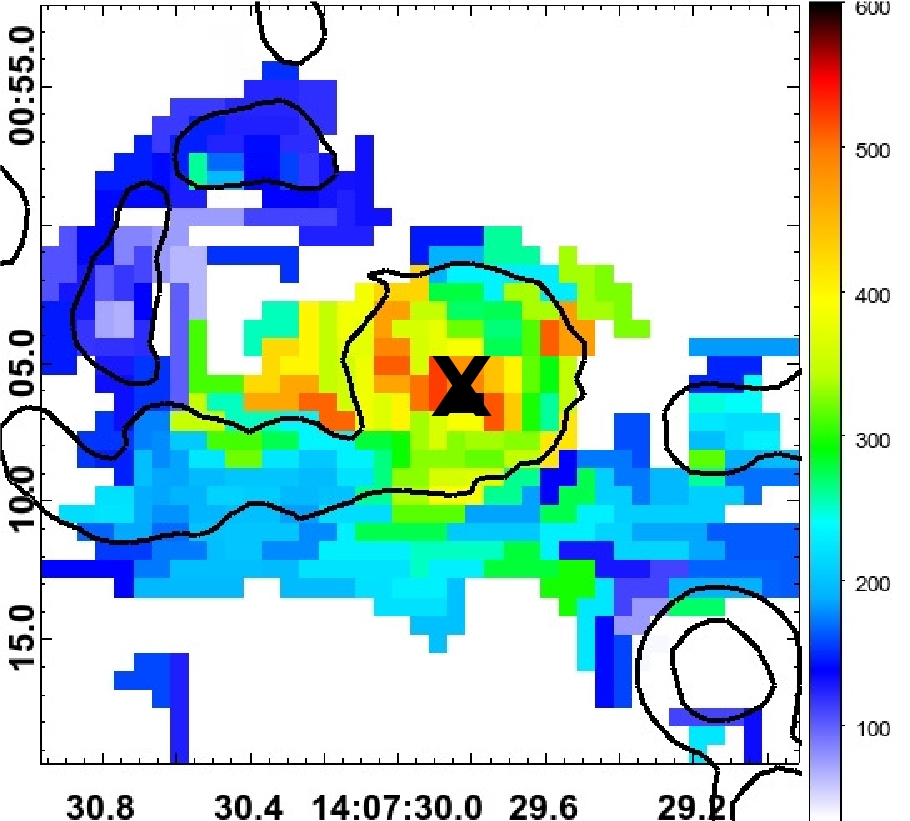}
\caption{The line-of-sight velocity of the narrow line component of the optical
nebulosity {\bf (left)} from a fit to the strong [N {\sc ii}]$\lambda$6583 line and
the FWHM velocity width {\bf (right)} after correction for the instrumental resolution
(98~\kmps, FWHM). Colour bars are in units of \kmps\ and the velocity zero point
is taken to be 6450~\kmps, the velocity of the H$\alpha$ nebulosity at the
position of the nucleus ($14^{h}07^{m}29^{s}.8$, $-27^{\circ}$01'06.4'', marked with an X). Only
pixels where the [N {\sc ii}]$\lambda$6583 emission line was detected above
4$\sigma$ are shown and the images are overlaid with contours of H$\alpha$ emission from our SOAR narrow-band images. \label{nebular_kinematics}}
\end{figure*}

The BGG in Abell 3581 has a smooth, relaxed appearance in broad band optical images.
However, the narrow band H$\alpha$+[N {\sc ii}] image shows striking filaments of
ionised gas. These filaments connect to the nucleus in the easterly and westerly
directions, parallel to the current radio axis, then curve sharply towards the
north and south respectively (see Fig. \ref{ha_image}). In the easterly direction the filaments arc round and apparently encase one of the previously detected X-ray bubbles. The filament system is
complex, formed of at least a few strands in each direction and is not one continuous
structure. Filaments of this nature have now been observed in many massive
galaxies dominating the centres of X-ray bright cool core groups and clusters.
Our VLT VIMOS IFS data show the emission lines of H$\alpha$, [N {\sc ii}], [S
{\sc ii}] and [O {\sc i}] in the inner filament all share the same morphology.

A dust lane stretching north to south through the nucleus is
clearly visible in all optical bands and in the F300X image. Fig.
\ref{dust_and_stars} shows the F555W image with the smooth elliptical component
subtracted, highlighting the myriad dust lanes. The nuclear component of the dust
lane is $\sim$5 arcsec in length and extends radially in two plume-like
structures; one to the north and one to the south. In the south, beyond a few arcsec, the dust lane abruptly changes direction and traces the
ionised gas filament to the east (see Fig. \ref{dust_and_halpha}). The dust
extends at least 10 arcsec along the filament; whether the dust stops
here or continues all the way along the filament is unknown. A similar extension
to the east is seen in the dust emission north of the nucleus. Contours of the dust
emission are overlaid on the net H$\alpha$+[N {\sc ii}] image in the left hand
panel of Fig. \ref{dust_and_halpha}. A young star
cluster is found coincident with the northern dust lane but no other obvious young 
stellar structures are observed in the filaments.

\begin{figure*}
\centering
\includegraphics[width=\textwidth]{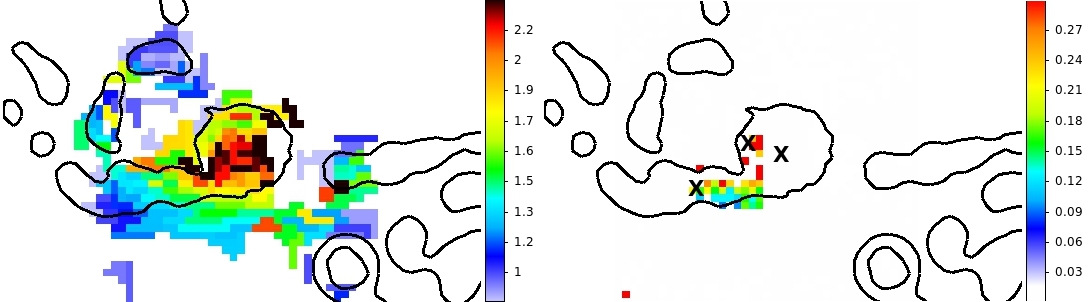}
\caption{{\bf Left:} [N {\sc ii}]$\lambda$6583/H$\alpha$ ratio map of the filaments
surrounding the two inner bubbles. The ratio is highest in the
centre of the galaxy and smoothly drops at larger radius. A spur of higher [N
{\sc ii}] is also seen north west of the centre where the
velocity dispersion in the gas is larger. {\bf Right:} The [O {\sc iii}]$\lambda$5007/H$\alpha$ ratio map on the same spatial and intensity scale as the right hand image. The ratios are corrected for Galactic extinction but no correction for internal extinction has been made. The three crosses indicate regions (the nuleus, NE and SE) where spectra were extracted and the emission line fluxes detailed in Table \ref{emission_lines} and the images are overlaid with contours of H$\alpha$ emission from our SOAR narrow-band images. \label{em_ratios}}
\end{figure*}

The right-hand panel of Fig. \ref{dust_and_halpha} shows an $E(B-V)$ map of the
relative intrinsic extinction in the central galaxy due to the dust lanes; the
reddening due to absorption by dust is significant in the central galaxy. This
is overlaid with contours of the JVLA L-band radio emission from the inner X-ray
cavities \citep{johnstone2005}. The dust lane in the south and extending towards the east of the
nucleus also coincides also with the outer edge of the radio bubble.

\subsubsection{Nebula kinematics}

Our VIMOS IFS data (Fig. \ref{spectrum}) provides evidence that a broad H$\alpha$ emission line is
necessary in the central 22 pixels, a region $\sim$2 arcsec in radius ($\sim$3 pixels or $\sim$1~kpc radius) and coinciding
with the largest surface brightness of high frequency radio emission. 
Our PSF is $\sim$2.1 pix FWHM so the emission should be resolved. The 
broad line region of the AGN is expected to be significantly smaller than this, being 
typically $<$0.1 pc or for the most luminous quasars $\sim$0.25 pc.
This broad component has
a best-fit line-of-sight velocity redshifted $\sim$150~\kmps\ with respect the
the narrow H$\alpha$ emission line component within the same region. The breadth
of the line is $\sim$2900~\kmps\ and the H$\alpha$
flux of the broad component in the inner 2 arcsec radius is
2.21$\pm$0.03$\times$10$^{-14}$\ergpcmsqps. 

The line-of-sight velocity and velocity width of the narrow component
of the ionised gas nebulae is shown in Fig. \ref{nebular_kinematics}. The units
of both figures are \kmps\ and the line-of-sight velocity zero point is taken to
be the nucleus ($14^{h}07^{m}29^{s}.8$, $-27^{\circ}$01'06.4'') as determined from the L-band JVLA radio images. Overlaid are
the contours from the narrowband H$\alpha$ image. A smooth line-of-sight velocity gradient is observed through the nucleus, perhaps indicative of rotation or an outflow. We note that no evidence for rotation of the stellar component, observed through the NaD absorption line, is seen. 

The gradient observed in the galaxy core extends
smoothly along the filament, and around the arc encasing the inner bubble, consistent with a common origin of the filament gas
and the gas in the core. The velocities are low in the filaments with
line-of-sight velocities of $\sim$50\kmps\ and velocity dispersions of
$\sim$150\kmps. The kinematics are therefore very similar to
those observed in filament systems surrounding other BCGs (e.g.
\citealt{hatch2005, canning2011b}).

\subsubsection{Emission line fluxes and ratios}

Our HRO data span the redshifted spectral range of [N {\sc i}]$\lambda$5199 to
[S {\sc ii}]$\lambda$6731 across the whole field-of-view and the redshifted spectral range of [O {\sc iii}]$\lambda$5007 to
[S {\sc ii}]$\lambda$6731 in the south-east and north-east quadrants. These data are fit within IDL as
described in Section \ref{ifu}. At the resolution of our observations, all
emission lines detected in our data have the same spatial morphology and, aside
from the nuclear region, our data are not of sufficient spectral resolution to constrain more than one velocity component. 

We check the fluxes in H$\alpha$ between our net H$\alpha$ image and our VIMOS IFU in a region on the inner south-east filament. The region is of size 19.5$\times$23 arcsec centred on $14^{h}07^{m}30^{s}.2$, $-27^{\circ}$01'04.8''. Our fluxes are 5.1$\times$10$^{-14}$~\ergpcmsqps\ and 5.7$\times$10$^{-14}$~\ergpcmsqps\ in the narrow-band image and IFU image respectively. The fluxes agree within $\sim$10 per cent. The error on the integrated flux from our {\sc idl} fit $\sim$ 7 per cent and a 1 per cent continuum flux variation between the ON/OFF filter images translates to $\sim$10 per cent net flux change for A3581. Our fluxes are therefore consistent.

The left hand plot of Fig. \ref{em_ratios} shows the ratio of the strong [N {\sc ii}]$\lambda$6583 emission line to the narrow component of the H$\alpha$ emission. In the centre of the galaxy this is observed to peak at the nucleus and to drop radially with distance from the AGN. Farther out onto the filaments the ratio is approximately unity. On the `inside' of the filaments, the side closest to the radio emission, a larger ratio of [N {\sc ii}] to H$\alpha$ emission is observed. 

The [O {\sc iii}]$\lambda$5007 emission line can only be observed in the eastern quadrants of the data cube; we have no information about the [O {\sc iii}]$\lambda$5007 emission in the nucleus (right hand panel of Fig. \ref{em_ratios}). The [O {\sc iii}]$\lambda$5007 emission is over three times stronger in the easterly extension to the north of the nucleus, where the velocity widths are very high, compared to the southern-most regions of the easterly filament. A gradient of [O {\sc iii}]$\lambda$5007 / H$\alpha$ emission, similar to that observed in the [N {\sc ii}]$\lambda$6583 / H$\alpha$ ratio discussed above, is observed across the more southern filament. The collisional line emission is enhanced relative to the recombination line emission on the `inside' of the filament and close to the nucleus.

Unusually strong emission lines of [N {\sc i}]$\lambda$5199, characteristic of the filament emission in BCGs, are detected in the extended filaments but no evidence is found for such emission in the nucleus (see Table. \ref{emission_lines}), we also detect He {\sc i} emission in the extended filaments. We do not observe any evidence for coronal emission lines (emission lines with excitation temperatures of $10^{5}-10^{6}$ degree gas) in our IFS data. Table. \ref{emission_lines} gives emission line ratios compared to H$\alpha$ for the three regions shown by crossed on Fig. \ref{em_ratios}. These are the nucleus, the north-easterly extension and farther out on the south-easterly extension. It is quite apparent that these ratios are different due to the comparative importance of various excitation mechanisms in these regions; this highlights the complicated nature of distinguishing excitation mechanisms in the central regions of BGGs/BCGs. 

\begin{table}
 \centering
 \caption{Emission line fluxes (upper panel) and ratios (lower panel) in the 3 regions indicated by a cross on Fig. \ref{em_ratios}. Each region was 2 pixels ($\sim$1.3 arcsec) in radius. The units for the fluxes are $\times$10$^{-15}$\ergpcmsqps.  $^{1}$Errors in nucleus large - probably due to multiple components. $^{2}$Spectral coverage of VIMOS quadrants is uneven so [O {\sc iii}]$\lambda$5007 emission can only be probed to the east of the nucleus. $^{3}$Broader line than H$\alpha$ due to not being able to resolve the doublet emission.}
 \begin{tabular}{c|c|c|c}
\hline
  Emission line & Nuc$^{1}$ & SE & NE \\
\hline \hline
  redshift & 0.021695$\pm$6E-6 & 0.021746$\pm$1E-6 & 0.02032$\pm$3E-5 \\
  velocity width & 301$\pm$6 & 209$\pm$1 & 355$\pm$36 \\
\hline
  {[}O {\sc iii}]$\lambda$5007 & N/A$^{2}$           & 0.4$\pm$0.02       & 0.4$\pm$0.02 \\
  {[}N {\sc i}]$\lambda$5199 & -             & 0.3$\pm$0.05$^{3}$ & 0.09$\pm$0.02$^{3}$ \\
  He~{\sc i}~5875 & -             & 0.1$\pm$0.02       & - \\
  {[}O {\sc i}]$\lambda$6300 & 1.7$\pm$0.4   & 0.7$\pm$0.01       & 0.2$\pm$0.03 \\
  {[}O {\sc i}]$\lambda$6363 & -             & 0.2$\pm$0.02       & - \\
  {[}N {\sc ii}]$\lambda$6548 & 2.7$\pm$0.1   & 1.5$\pm$0.04       & 0.3$\pm$0.01 \\
  H$\alpha$~6563 & 4.6$\pm$0.1   & 3.1$\pm$0.01       & 0.6$\pm$0.03 \\
  {[}N {\sc ii}]$\lambda$6583 & 11.4$\pm$0.07 & 4.5$\pm$0.02       & 1.1$\pm$0.03 \\
  {[}S {\sc ii}]$\lambda$6717 & 3.2$\pm$0.2   & 1.5$\pm$0.02       & 0.6$\pm$0.02 \\
  {[}S {\sc ii}]$\lambda$6731 & 3.0$\pm$0.2   & 1.2$\pm$0.03       & 0.4$\pm$0.1  \\ 
\hline
  Line / H$\alpha$. & & \\
\hline \hline
  {[}O {\sc iii}]$\lambda$5007 & N/A$^{2}$  & 0.13   & 0.67 \\
  {[}N {\sc i}]$\lambda$5199 & -    & 0.10   & 0.15 \\
  He~{\sc i}~5875 & -    & 0.03   & - \\
  {[}O {\sc i}]$\lambda$6300 & 0.37 & 0.23   & 0.33 \\
  {[}O {\sc i}]$\lambda$6363 & -    & 0.06   & - \\
  {[}N {\sc ii}]$\lambda$6548 & 0.59 & 0.48   & 0.5 \\
  H$\alpha$~6563 & 1.0  & 1.0    & 1.0 \\
  {[}N {\sc ii}]$\lambda$6583 & 2.48 & 1.45   & 1.83 \\
  {[}S {\sc ii}]$\lambda$6717 & 0.70 & 0.48   & 1.0 \\
  {[}S {\sc ii}]$\lambda$6731 & 0.65 & 0.39   & 0.67  \\ 
 \end{tabular}
 \label{emission_lines}
\end{table}

\subsubsection{Star formation}

A young star cluster is observed, projected on the dust lane, 2.5 arcsec north of the nucleus (see Fig. \ref{dust_and_stars}). No other young clusters are observed in the filament system. Using our relative reddening map, and assuming the intrinsic dust is an obscuring screen, we correct the observed fluxes in the broadband filters and determine the emission from the young star cluster. Using a {\sc galev} evolutionary synthesis model \citep{kotulla2009} for a simple stellar population the F300X-F555W and F300X-F814W colours indicate an age $<$10$^{8}$~yrs and a mass $\sim$10$^{4}$ \Msun. 

Abell 3581 has not been observed by Spitzer. However, the central galaxy is detected in all four WISE bands. The WISE All-Sky Source Catalog, measures an extrapolated flux at 24 $\mu$m as 8.0 mJy. Using equation 10 of \cite{rieke2009}, and assuming a \cite{kroupa2002} IMF, indicates the star formation rate (SFR) in the central galaxy is $\sim$ 0.21 \Msunpyr. The near- and far-ultraviolet emission from Galex indicates 0.3~\Msunpyr and 0.1~\Msunpyr\ (these values assume no intrinsic reddening correction).

The net H$\alpha$ image of IC~4374 is shown in Fig. 1. The total H$\alpha$ flux is 6.9$\times10^{-14}$ erg cm$^{-2}$ s$^{-1}$, assuming [N {\sc ii}]6583/H$\alpha$ = 1.5 and [N {\sc ii}]6548/[NII]6583 = 1/3 (an average from our VIMOS IFS data - we note this does not cover the entire extent of the filament system). If the nucleus is excluded (a circle with a radius of 2.5 arcsec), the total flux is 4.8$\times10^{-14}$ erg cm$^{-2}$ s$^{-1}$. Using the \cite{kennicutt1998} relation, SFR (\Msunpyr) = 7.9$\times$10$^{−42}$ (L$_{\mathrm{H\alpha}}$ / ergs s$^{-1}$), the SFR is $\sim$ 0.4~\Msunpyr, assuming no intrinsic extinction for the H$\alpha$ emission. This is similar to the mass deposition rate in the cool X-ray gas (0.4~\Msunpyr, \citealt{sanders2010}), however, no UV emission is observed in the filaments.

\subsection{X-ray}
\label{xray_section}

\subsubsection{Morphology}

\begin{figure*}
\centering
\includegraphics[width=0.85\textwidth]{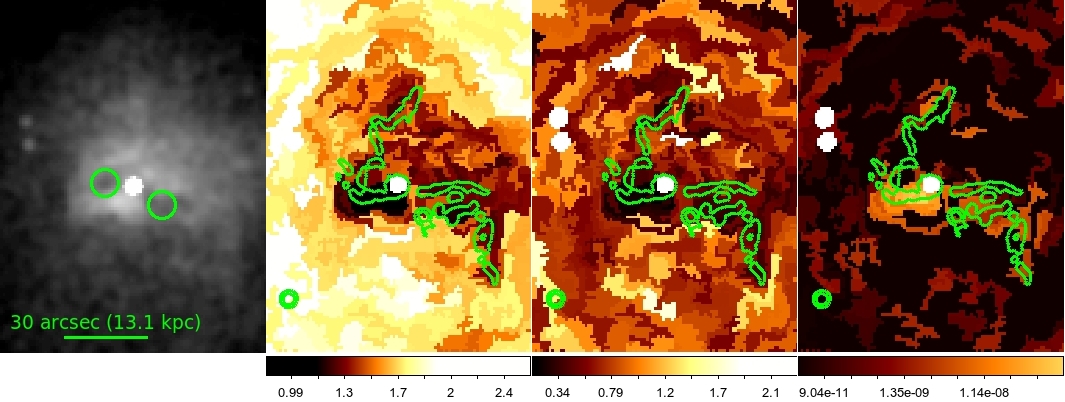}
\caption{{\bf From left:} X-ray surface brightness image, X-ray temperature map, X-ray
metallicity map and 0.5~keV emission measure map. The inner two depressions
corresponding to the regions in which 1.4~GHz radio emission is seen are circled
on the X-ray surface brightness image. Contours of H$\alpha$ emission are
overlaid on the temperature, metallicity and emission measure maps. The colour bars are in units of keV, Z$_{\odot}$ and \empasecsq\ respectively.
\label{xray_optical}}
\end{figure*}

\begin{figure*}
\centering
\includegraphics[width=0.85\textwidth]{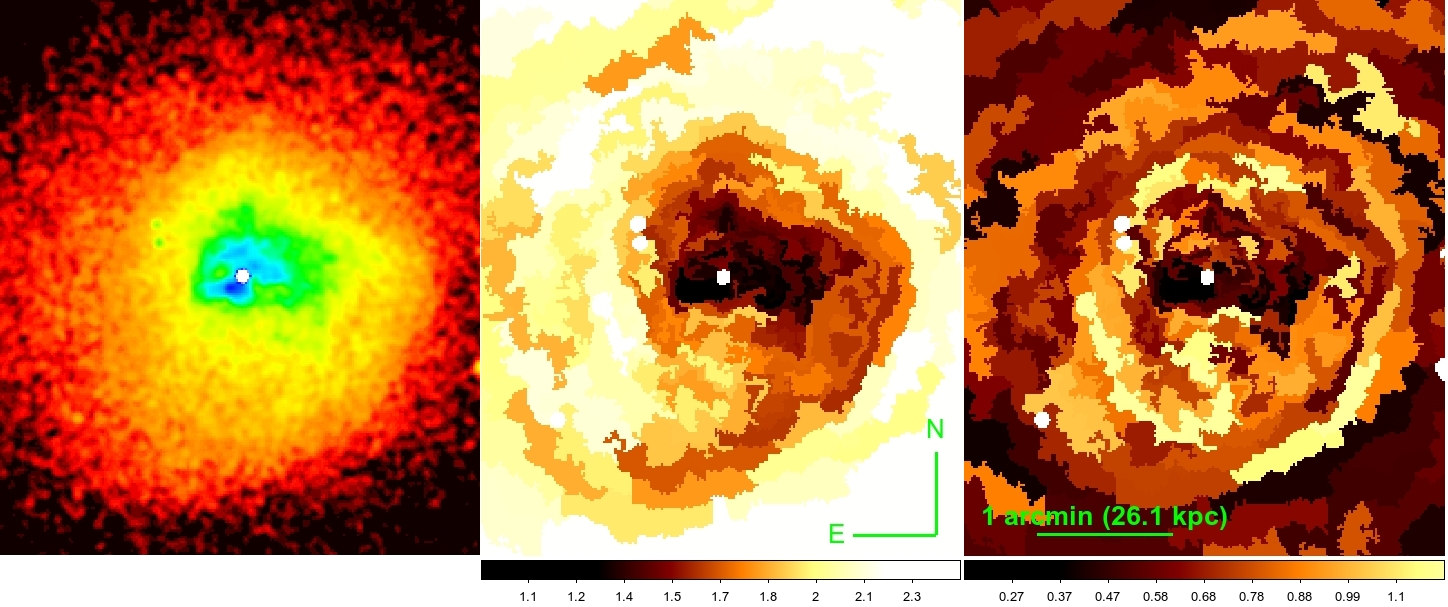}
\caption{{\bf From left:} X-ray surface brightness image, X-ray temperature map and X-ray
metallicity map. Two depressions are observed which
coincide with the 1.4~GHz radio emission in the nucleus, an outer cavity is observed towards the south west beyond a sharp temperature and metallicity edge and a further surface brightness discontinuity, a likely cold front, is observed beyond this outer bubble. There is an apparent metallicity drop toward the centre, as seen in other
similar objects, this could indicate the need for additional
temperature components to the model (see discussion in section \ref{xray_abund}). \label{xray_outer}}
\end{figure*}

\begin{figure*}
\centering
\includegraphics[width=0.8\textwidth]{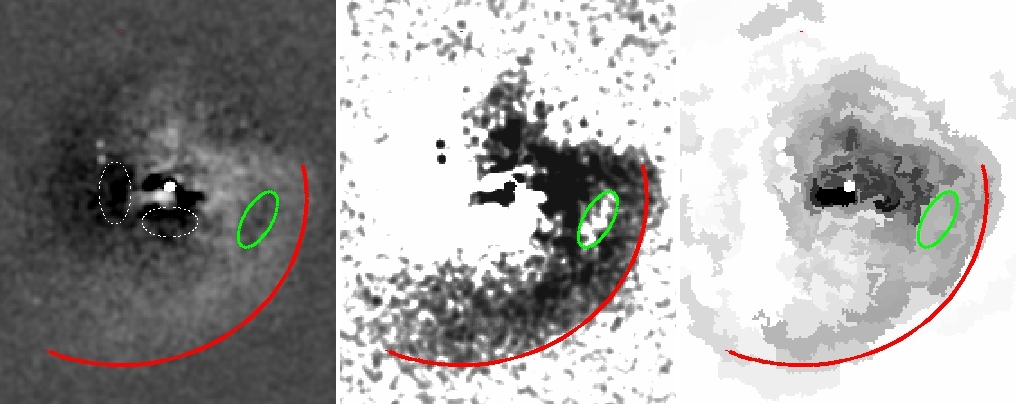}
\caption{The outer features. {\bf Left:}  Unsharpmasked {\it Chandra} X-ray image of  Abell 3581. An image, smoothed with a gaussian with $\sigma=$25 pixels has been subtracted from the original. The outer cavity is highlighted in green and the cold front in red. The two `bays' to the east and south are indicated with dashed white ellipses. {\bf Middle:} The X-ray surface brightness image after a smooth elliptical isophotal model was subtracted from the original data. The intensity scale was chosen to highlight the outer cavity and spiral. {\bf Right:} The same regions overlaid on the X-ray temperature map. \label{xray_features}}
\end{figure*}

The X-ray surface brightness, temperature and
metallicity maps are shown in the core in Fig. \ref{xray_optical} and on a larger
scale in Fig. \ref{xray_outer}. New outer structures are indicated in Fig.
\ref{xray_features}. As previously noted by \cite{johnstone2005}, who examined a
7165~s \textit{Chandra} ACIS-S observation of Abell 3581, an X-ray emitting
point source lies in the nucleus and two inner cavities exist; one to the east
and one to the west of the nucleus. These inner cavities coincide with
relativistic synchrotron emitting gas observed as radio bubbles in 1.4~\GHz\
radio emission. The power in these inner 1.4 GHz radio
bubbles is 2.5$\times$10$^{23}$~W~Hz$^{-1}$ \citep{johnstone2005}.

Fig. \ref{xray_optical} reveals that the low volume filling, soft X-ray gas, detected by \cite{sanders2010}, lies in 
filaments. There is excellent spatial correspondence between
the coolest, highest density X-ray emitting gas and the H$\alpha$ emitting filaments. Clear structures are seen in the temperature and
metallicity maps suggesting little mixing is occurring in the centre of Abell
3581. 

Additional structure is observed outside of the filaments; surface brightness edges to the east and south indicate two bays where a dearth of X-ray emission could be a signature of outer bubbles (see Fig. \ref{xray_features}, white ellipses). An additional X-ray cavity (green ellipse), detached from the nucleus and not observed before, is seen $\sim$45 arcsec to the
west of the nucleus. We do not observe any radio emission, in our frequency bands, coincident with
the depression in the X-ray image. The cool X-ray gas is observed to extend up to the new outer bubble but not farther. A clear edge
is seen in both maps at the inner edge of the bubble.

\begin{figure}
\centering
\includegraphics[width=0.4\textwidth]{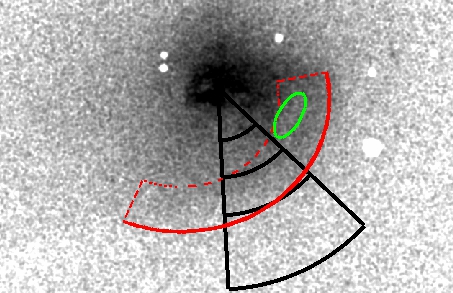}
\includegraphics[width=0.4\textwidth]{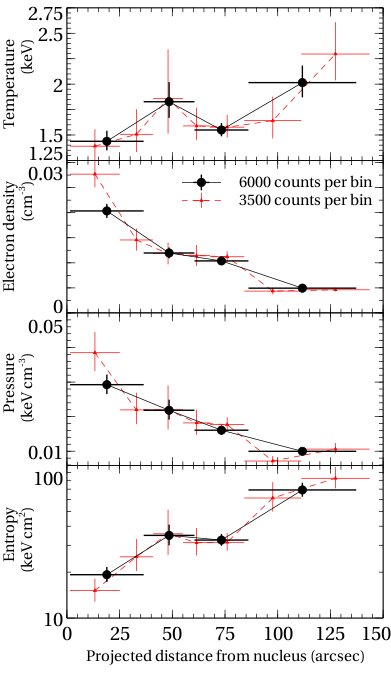}
\caption{Deprojected profiles showing the temperature, density, pressure and entropy of a sector centred on the cool core and intersecting the spiral. The entropy and temperature of the gas in the spiral is lower than that of the gas between the spiral and the core. The black regions contain 6000 counts and are shown on the above plot with the outer bubble, spiral cold front and approximate width of the spiral also shown. The red regions contain 3500 counts such that the statistics remain good while allowing two regions on the spiral. \label{spiral_ent}}
\end{figure}

The large scale structure in the X-ray image shows a spiral surface brightness
discontinuity (red curve) which wraps around the nucleus in a clockwise direction. The X-ray gas is cooler, denser and
higher metallicity within this structure and the entropy, defined as $K=k_{B}T/n_{e}^{2/3}$, is lower than the surrounding gas, similar to the gas in the X-ray cool core and may be gas stripped from there by sloshing motions in the core. Fig. \ref{spiral_ent} shows the X-ray properties through a sector centred on the nucleus and intersecting part of the spiral structure.

\subsubsection{X-ray metallicity}
\label{xray_abund}

The X-ray metallicity maps (Fig. \ref{xray_optical} and Fig. \ref{xray_outer}) show a high abundance spiral structure but a low abundance core to the galaxy. Most cool-core galaxy clusters and groups tend to show a central abundance peak (e.g. \citealt{degrandi2001}). However, an offset peak and metal enriched spiral is also observed in the Centaurus cluster \citep{sanders2002}, Ophiuchus Cluster \citep{million2010} and HCG 62 (e.g. \citealt{gu2007, gitti2010, rafferty2012}). HCG 62 has a central iron abundance of $\sim$0.3~$Z_{\odot}$ which rises to solar at 20-30 arcsec ($5.6-8.4$~kpc) before dropping father out \citep{rafferty2012} while Centaurus has a central iron abundance of $\sim$0.4~$Z_{\odot}$ rising to $\sim$1.3~$Z_{\odot}$ 15~kpc from the nucleus \citep{sanders2002}. Similarly, with a single temperature fit to the data we find an iron abundance of $\sim$0.3~$Z_{\odot}$ in the central regions of Abell 3581 which rises to 0.8-1.0~$Z_{\odot}$ at a distance of $\sim30-60$ arcsec ($13.5-26$~kpc) coinciding with the spiral structure. The metallicity then drops farther out into the ICM. 

Resonance scattering can effect the measured abundances in clusters and groups; however, the effect has not been able to fully explain the metallicity drops (e.g. \citealt{mathews2001, gastaldello2004, sanders2006b}). A discussion on the effect of LMXBs on the derived spectral properties is given in \cite{david2009,david2011}. The gas in groups and clusters, viewed in projection, is certainly multi-phase and another possibility for the metallicity drop in the central regions of the galaxy may be that a single temperature model is inadequate in representing the data.

\begin{figure}
\centering
\includegraphics[width=0.4\textwidth]{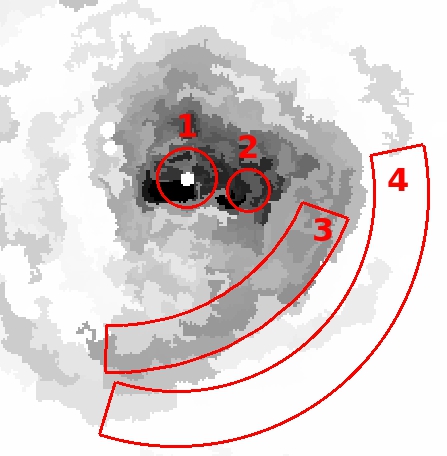}
\caption{Regions 1-4 where the X-ray spectra has been extracted and a multi-temperature gas is fit to the data. The point source in region 1 has been removed. The results of the single temperature and two temperature fits are given in Table \ref{fitting}. \label{reg_spec}}
\end{figure}

To examine whether the metallicity drop is due to the misrepresentation of a multi-phase gas, we fit two temperature {\sc vapec} models within Xspec (version 12.8.0a) to the X-ray emission in four regions within the cluster. These regions are shown in Fig. \ref{reg_spec} and the results of the fit are given in Table \ref{fitting}. The minimum counts in our regions are 5400 counts with 2 of the regions having greater than 12000 counts. The results of Table \ref{fitting} show that the addition of a second temperature component in regions 1 and 2 significantly alters the abundances removing any evidence for a metallicity drop in the central regions. The spiral (region 3) has a similar best fit metallicity as the core while beyond the spiral (region 4) the metallicity declines.

\begin{table*}
 \centering\
 \caption{Results from the multi-temperature fits to regions 1-4 in Fig. \ref{reg_spec}. }
 \begin{tabular}{c|c|c|c|c|c|c|c|c}
  Region & Model & C stat & $kT$ & Norm & Si  & S  & Fe & Ni \\
         &       &                  & (keV) & (cm$^{-5}$) & \multicolumn{4}{c}{Abundances in solar units}    \\
\hline \hline
  Region1 & {\sc vapec} & 336.94/315 & $1.29_{-0.02}^{0.02}$ & $0.00088_{-6.6e-05}^{6.7e-05}$ & $0.58_{-0.09}^{0.10}$ & $0.65_{-0.16}^{0.18}$ & $0.27_{-0.05}^{0.05}$ &  $0.62_{-0.41}^{0.44}$ \\
         &  & 1.070 &  &  &  &  & & \\
   & 2$\times${\sc vapec} & 316.743/313 & $ 0.36_{-0.08}^{ 0.04}$ & $0.00011_{-2.1e-05}^{5.9e-05}$ & $ 1.11_{-0.29}^{ 0.35}$ & $ 1.04_{-0.31}^{ 0.37}$ & $ 0.40_{-0.09}^{ 0.11}$ & $ 1.45_{-0.77}^{ 0.82}$ \\
         &  & 1.011 & $ 1.50_{-0.08}^{ 0.08}$ & $0.00057_{-8.6e-05}^{0.00011}$ &  &  &  &   \\
\hline 
  Region2 & {\sc vapec} & 229.293/251 & $ 1.32_{-0.04}^{ 0.03}$ & $0.00029_{-3.6e-05}^{2.2e-05}$ & $0.68_{-0.16}^{0.21}$ & $0.86_{-0.29}^{0.37}$ & $0.33_{-0.087}^{0.08}$ & $0.86_{-0.79}^{0.95}$ \\
         &  & 0.914 &  &  &  &  &  &   \\
   & 2$\times${\sc vapec} & 220.801/249 & $1.5_{-0.12}^{0.16}$ & $0.00021_{-5.8e-05}^{5e-05}$ & $1_{-0.29}^{0.58}$ & $1.1_{-0.41}^{0.69}$ & $0.51_{-0.16}^{0.27}$ & $0.99_{-0.99}^{1.4}$ \\
         &  & 0.887 & $0.81_{-0.17}^{0.22}$ & $1.9e-05_{-8.1e-06}^{2.4e-05}$ &  &  &  &   \\
\hline
  Region3 & {\sc vapec} & 335.605/340 & $1.6_{-0.054}^{0.045}$ & $0.00055_{-3.3e-05}^{3.4e-05}$ & $1_{-0.17}^{0.18}$ & $0.73_{-0.25}^{0.27}$ & $0.48_{-0.086}^{0.08}$ & $1.2_{-0.66}^{0.7}$ \\
         &  & 0.987 &  &  &  &  &  &  \\
   & 2$\times${\sc vapec} & 332.939/338 & $17_{-17}^{-17}$ & $1.8e-05_{-1.8e-05}^{0.00049}$ & $1.2_{-0.24}^{0.31}$ & $0.85_{-0.29}^{0.36}$ & $0.5_{-0.11}^{0.12}$ & $1.8_{-0.95}^{1.3}$ \\
         &  & 0.985 & $1.6_{-0.33}^{0.063}$ & $0.00049_{-0.00049}^{7.3e-05}$ &  &  &  &  \\
\hline
  Region4 & {\sc vapec} & 447.967/408 & $1.99_{-0.25}^{0.23}$ & $0.0004566_{-0.0000687}^{0.0000605}$ & $0.42_{-0.29}^{0.32}$ & $0.43_{-0.43}^{0.51}$ & $0.35_{-0.13}^{0.16}$ &  $0.62_{-0.62}^{1.10}$ \\
         &  & 1.09 &  &  &  &  &  &  \\
   & 2$\times${\sc vapec} & 447.966/406 & $0.08_{-0.08}^{-0.08}$ & $0.00_{0.822}^{-0.00}$ & $0.42_{-0.29}^{0.32}$ & $0.44_{-0.44}^{0.50}$&$0.34_{-0.13}^{0.17}$ & $0.63_{-0.63}^{1.09}$ \\
         &  & 1.10 & $1.98_{-0.24}^{0.23}$ & $0.00045_{-0.000069}^{0.000062}$ &  &  &  &   \\
 \end{tabular}
 \label{fitting}
\end{table*}

\begin{figure}
\centering
\includegraphics[width=0.4\textwidth]{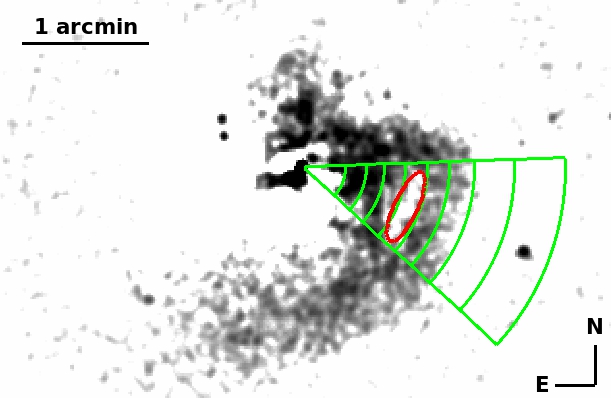}
\includegraphics[width=0.4\textwidth]{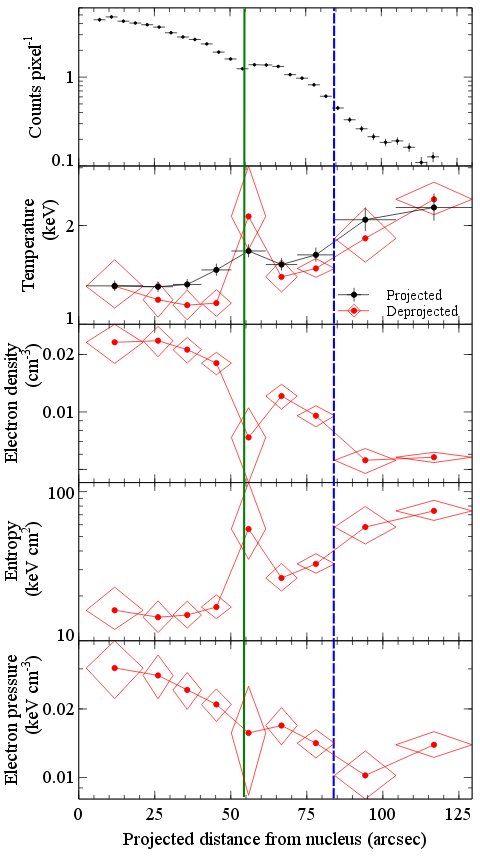}
\caption{Temperature, density and pressure profiles in a sector chosen to
include the cavity and cold front. The widths of the annuli are determined by
requiring a minimum of 3000 counts per annuli and are shown in the image above. 
The green solid line on the figure indicates the radius of the center of the outer cavity 
while the blue dashed line indicates the radius of the cold front. \label{cavity_profile}}
\end{figure}

\subsubsection{Bubble energetics}

To determine the energetics of the new outer cavity we fit an elliptical isophotal model to the
cluster and subtract the median surface brightness at each radius. We then
identify, by examining cuts through the resulting surface brightness image, 
an ellipse representing the cavity.

We define a sector of the cluster centred on the nucleus and extending between
the edges of the ellipse (see Fig. \ref{cavity_profile}) up to a radius of
100 arcsec (44~\kpc). We create annular regions within this sector. The width of
each annulus is determined to be the width necessary to enclose at least 3000
counts; we do this to ensure a good fit to the spectrum. The inner 3 pixels are
not used to avoid including the point source. We then fit both
the projected and deprojected spectra within Xspec (as described in
Section \ref{x-ray_data}).

Estimating errors on energetics from cavities is fraught with
difficulties due to inferring a 3 dimensional shape for the bubble from a
projected 2 dimensional image, not knowing the inclination of the rise direction
to the observer, the `fuzzy' edges of the bubble decrement and due 
to the shape of the overlying surface brightness profile
of the cluster. 
The errors estimated and quoted in the following sections use the technique of \cite{osullivan2011}. Using this technique the line-of-sight depth is equal to the semi-minor axis and the uncertainty on the volume is given by assuming a minimum depth of half this value and a maximum equal to the semi-major axis. However, for the outer cavity, it must be noted that the length of the bubble in a direction perpendicular to the rise direction is given by the semi-major, not semi-minor axis of the ellipse; the errors quoted are unlikely to be reliable.  

The temperature, density, pressure and entropy profiles in the gas in this sector are
shown in Fig. \ref{cavity_profile}. At the position of the X-ray cavity
(indicated by a green solid line) we detect a clear jump in the deprojected temperature and a corresponding drop in
the density. There is no significant change in the
pressure which could indicate that the jump in temperature is due
to the gas in the region around the bubble and not the outer layers of
the cluster seen in projection with the cavity. However, the errors on the pressure are not constraining.

We estimate the energy in the outer cavity using,
\begin{equation}
 E_{bubble}=\frac{\gamma}{\gamma - 1}\mathrm{P}V,
\end{equation}
where $\gamma=4/3$ for a relativistic fluid, $V$ is the volume of the cavity and
$\mathrm{P}$ is the thermal pressure of the ICM, as described in
\cite{birzan2004} and \cite{dunn2005}. The volume of the bubble is derived making the
implicit assumption that the bubble is a spheroid, two of whose axes are in the plane of the
sky, with a plane of symmetry perpendicular to the rise direction. The ellipsoid
volume $V=(4/3)\pi R_{1}R_{2}^{2}$, where $R_{1}=3.8$\kpc\ (8.8\arcsec) is the
length of the projected ellipsoid axis in the direction of the nucleus, in this
case the semi-minor axis of the ellipse, and $R_{2}=8.4$\kpc\ (19.3\arcsec) is
the length of the ellipsoid in a perpendicular direction, here given by the
semi-major axis of the ellipse. Using this formalism,
$E_{bubble}=5.5 (2.7-11.8\times10^{56})$~ergs. A similar treatment gives the energy
in the inner cavities as $E_{bubble}=2.4(1.2-2.9\times10^{56})$~ergs, assuming
$R_{1}=3.6$\kpc\ (8.4\arcsec) and $R_{2}
=3.0$\kpc\ (6.9\arcsec). \cite{diehl2008} estimate similar
energies for these inner bubbles with values of 3.49$\times$10$^{56}$ and 3.37$\times$10$^{56}$ for the inner bubble energies.

The outer bubble rise time as measured by the sound speed, buoyancy and refilling timescales (\citealt{blanton2001, mcnamara2000,
churazov2000}) are 3, 4, and 5$\times10^{7}$~years respectively and for the inner bubbles are and 0.7, 2 and 6$\times10^{7}$~years. Assuming the sound speed timescale for the inner bubbles and the buoyancy timescale for the outer bubble the cavity powers are $P_{x-cav}=4.4\times10^{41}$~ergs~s$^{-1}$ for the outer cavity and 
$P_{x-cav}=1.1\times10^{42}$~ergs~s$^{-1}$ for the inner cavities.
A simple scaling from the power in the 1.4~GHz radio emission (2.5$\times$10$^{23}$~W~Hz$^{-1}$), using the
relation derived by \cite{cavagnolo2010}, 
gives $P_{cav}=3.7\times10^{43}$~ergs~s$^{-1}$ for the inner cavities.

The point source removed X-ray luminosity, between 0.5-7~keV, in a circle of radius 9.4~kpc (21.5'') which encompasses the inner bubbles, a circle of radius 28.3~kpc (65.0'') which encompasses the outer bubble and the `cooling radius' (assumed to be where $t_{cool}<7.7$~Gyr, $r_{cool}\sim57$~kpc, \citealt{sanders2010}), corrected for Galactic absorption are 2.1$\times$10$^{42}$, 7.8$\times$10$^{42}$, and 1.1$\times$10$^{43}$~\ergps\ respectively.

\subsection{Radio}

\subsubsection{Morphology}

The morphology of the radio emission is shown in Fig. \ref{xray_radio}. The high
frequency 1.4 GHz emission fills the inner cavities; these are surrounded
by the ionised gas nebulae which coincides with the lowest temperature X-ray
emission. This is consistent with the inflation of the bubbles leading to the
displacement of the cool and warm ISM in the galaxy, due to the pressure of the
relativistic particles.

The left hand panel of Fig. \ref{xray_radio} compares the morphology of the lower frequency radio emission
to that of the high frequency radio emission and the X-ray gas. At lower
frequencies the orientation of the radio extension is rotated clockwise with respect to the high frequency emission. The rotation is in the same sense as the H$\alpha$ filament (see Fig. \ref{xray_radio}, right hand panel) and the 614~MHz contours (green contours) show a spur of emission which follows the filament in the north-easterly direction. 

\section{Discussion}

A distinguishing feature of many BCGs in cool core clusters, where heating is required
to prevent catastrophic cooling of the hot gas, is that they harbour intricate extended
systems of low ionisation optically emitting gas. A significant body of work has
studied the origin of these filaments and their relationship with star
formation, AGN feedback and gas in other phases. The BGG IC 4374 in Abell 3581 is one such system
where it is clear that the structures in the hot, cool and cold dusty gas are
intimately related and are connected with the presence of the radio lobes.

IC 4374 exhibits prominent dust lanes running north and south through the galaxy nucleus. Both these dust lanes have extensions to the east which trace the ionised gas
filaments (Fig. \ref{dust_and_stars}). A young star cluster is observed in the
northern wedge shaped dust lane but no other evidence for star formation,
notably none in the filament system, is observed.

Scaling from our $E(B-V)$ map (Fig. \ref{dust_and_halpha}) of relative intrinsic
dust extinction and assuming a hydrogen column density of
N$_{\mathrm{H}}=5.8\times10^{21} \times E(B-V)$ and the Galactic gas to dust
ratio, $\sim$100, allows us to estimate the dust mass in the galaxy as
10$^{6}$~\Msun\ with $\sim$10$^{5}$\Msun\ in the south-east filament.

\begin{figure*}
\centering
\includegraphics[width=0.4\textwidth]{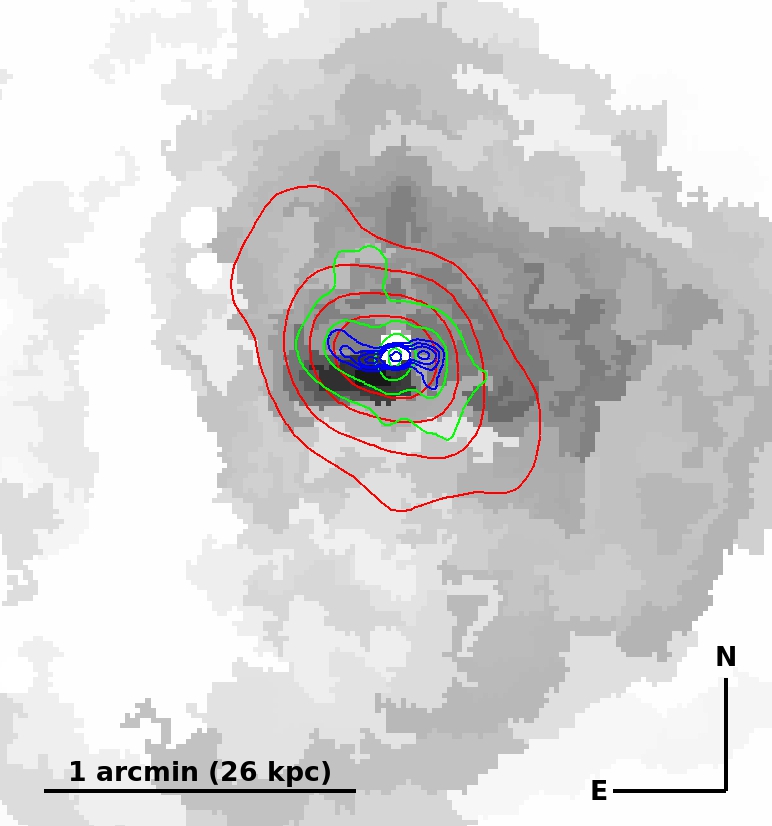}
\includegraphics[width=0.4\textwidth]{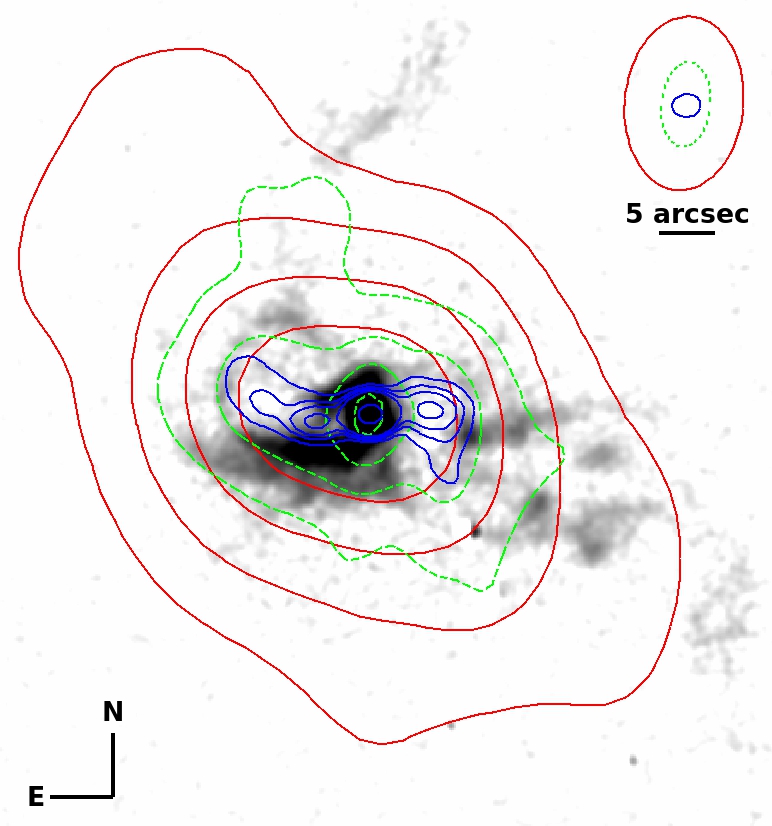}
\caption{{\bf Left:} X-ray temperature map overlaid with contours of radio emission. Red
contours are from 244~MHz GMRT emission, green contours from 614~MHz GMRT emission and the
central blue contours are from L-band JVLA radio emission. Low temperature gas (darker colours, see Fig. \ref{xray_optical} and \ref{xray_outer}) is observed at the centre of the system and stretches east and west from the nucleus towards the bubbles. {\bf Right:} The same radio contours overlaid on the {\sc soar} net H$\alpha$ image. The beam sizes are shown in the top right hand corner of the image. \label{xray_radio}}
\end{figure*}

The morphology of the H$\alpha+$[N {\sc ii}] emission is matched very well by
the coolest and highest density
regions of the X-ray gas (Fig. \ref{xray_optical}). The optical filaments engulf the inner bubbles and are coincident with the easterly extending dust lanes. The smooth velocity gradients of
the ionised gas suggest the filaments extending around the bubbles are
continuous. 

The south-west filament leads to an outer bubble. Cool, low
metallicity X-ray gas also extends from the centre of the cluster in the
direction of this bubble and there is a discontinuity in the properties of the X-ray gas at
a radius corresponding to the inner edge of the bubble. The lower frequency
radio emission is rotated clockwise with respect to the high frequency emission
and is aligned along the axis of the outer, not the inner, bubble (see Fig. \ref{xray_radio}). 

A spiral of high metallicity, low temperature, low entropy X-ray gas, with a cold front at its outer edge, is observed to trail from the cool core. In the following sections we draw on these observations to attempt to answer questions pertinent to the nature of the heating and cooling cycle in this system.

\subsection{Origin of filaments and dust lanes}

The filaments in Abell 3581 exhibit point symmetry but are not radial, varying direction as they
extend out of the galaxy. We cannot comment on the extent of the directional
change as this depends on the projection angle. The velocity structure of the filaments is smooth, consistent with the cool gas being pushed aside due to the expansion of the inner radio bubbles and, importantly, both the eastern filament extensions (the blueshifted and redshifted sides of the filament) are accompanied by thin threads of dust. 

Initial dust grain formation requires high densities and low enough temperatures to permit the condensation of silicates and graphites; it therefore cannot form straight out of a dust-free ICM. However, \cite{draine2003} makes the point that our current knowledge of grain destruction time-scales and observed element depletions in the gas-phase highlight that much of the subsequent dust growth must happen through interstellar processing. Dust and polycylic aromatic hydrocarbon (PAH) features have both been found in the extended ionised gas filaments of other BCGs (e.g. \citealt{sparks1989, johnstone2007, donahue2011}).

The rim of the inner bubble, east of the nucleus, is laced with emission from ionised gas apparently encasing the bubble, implying the hot gas inflating the bubble is sweeping aside the cooler ionised gas. The south-west filament ends at the discontinuity in surface brightness, temperature and metallicity delineating the inner edge of the outer bubble indicating this filament may have been dragged up beneath the outer bubble. The correlation
observed between the cool gas and dust filaments, X-ray cavities and the radio emission and the symmetry of the
two filament extensions is hard to reconcile within the framework of filaments
cooling from the hot ICM in situ or being the result of a galaxy merger.

\begin{figure}
\centering
\includegraphics[width=0.45\textwidth]{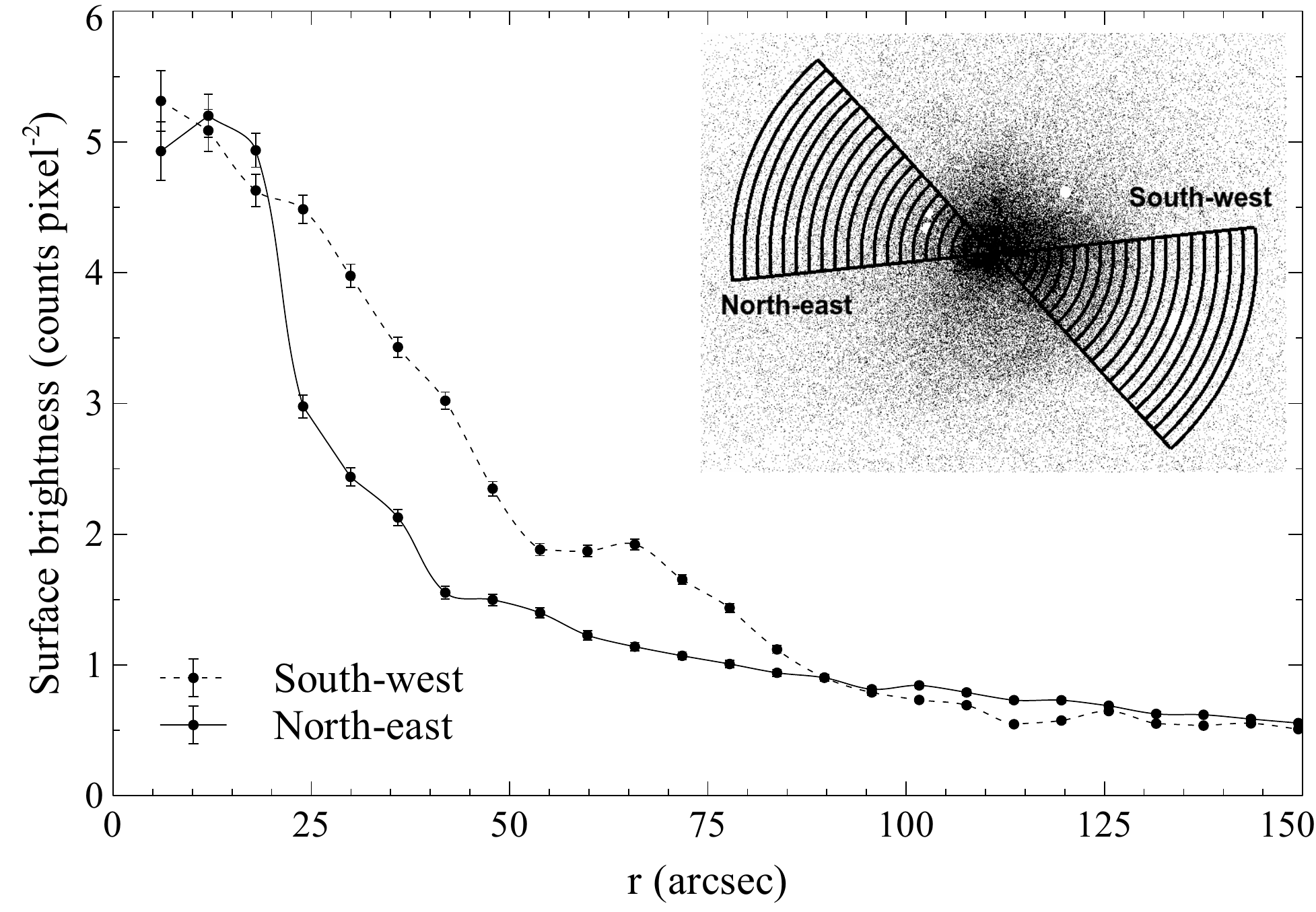}
\caption{X-ray surface brightness profile towards the outer X-ray cavity
(south-west) and in a direction reflected through the nucleus (north-east). The
X-ray emission is not radially symmetric; in the north-east direction the
cluster surface brightness drops much more sharply than towards the south-west.
The outer cavity in the south-west is at
a distance of $\sim$55 arcsec. \label{sb_profile}}
\end{figure}

It is therefore tempting to advocate that this outer bubble and another in the opposite
direction are responsible for the displacement of cool gas seeding the extended filaments and an interaction between the cool uplifted material and the hot ICM is responsible for the cooler X-ray gas (e.g. see discussion in \citealt{werner2012}) However, we note that by eye, we do not observe a second outer bubble. The X-ray surface
brightness is not symmetrical about the nucleus; the surface brightness drops more sharply in the
north-east of Abell 3581 than in the south-west (Fig. \ref{sb_profile}) which might inhibit our view of a second outer bubble.

Crudely, we test the hypothesis that the corresponding north-eastern outer bubble is
not observed due to the low surface brightness by examining the significance of
a similar outer bubble (same size/shape), reflected through the nucleus of Abell 3581.

To quantify the significance of a bubble we compare the counts in the `bubble region' to regions on either side of the bubble and with the same shape and radius. For our detected outer bubble we find a $\sim$10 per cent decrement in counts ($\sim$3010 counts compared with $\sim$3390 counts in the undisturbed regions) corresponding to a $\sim$6 $\sigma$ deficit. If we assume a similar size and shape bubble with point symmetry through the nucleus, we find a $\sim$5 per cent decrement ($\sim$2119 and $\sim$2196 counts) corresponding to $\sim$1.5 sigma. A 10 per cent deficit in counts in this low surface brightness region, similar to the deficit seen in the outer bubble, would still only amount to a 3 sigma detection of a broad bubble and is probably not easily observed by eye. We may not expect to see a bubble with the same size and shape and at the same radius. The surface brightness drop indicates a lower gas density. We might expect a bubble to be larger in size in a less dense medium further decreasing the likelihood of observing it. We cannot rule out the possibility of another outer bubble being responsible for the gas uplift of the filaments towards the north-east. We also note that a depression in the surface brightness profile is observed at $\sim$40 arcsec from the nucleus towards the north-east direction (see Fig. \ref{sb_profile}) which lends supporting evidence for another outer bubble.

The emission lines in the central regions of the galaxy have very large measured velocity dispersions particularly
towards the eastern inner cavity. Our data are unable to constrain two or more velocity components to the line emission but it is likely that multiple velocity components contribute to the broad velocity features observed.
\cite{farage2012} suggest the broad blueshifted feature is indicative of an inflow, perhaps responsible for feeding the AGN.
We suggest the line-of-sight velocities, blueshifted in one direction and 
redshifted in the other, are consistent with the expansion of a radio bubble
pushing aside the cool gas as it expands. This cool gas may have existed in an extended structure due to outflows from previous bubbles. The action of the current inner radio bubbles is to sweep aside the cool gas and to expand in the direction of least resistance. The coincidence of both the north-east and south-east filaments with filamentary dust lanes supports the idea that the gas is being lifted/pushed away from the centre of the galaxy.

High spatial resolution observations of the cold gas, for example with ALMA or SINFONI IFU, would shed light on the presence of an inflow or outflow in the nucleus.

\subsubsection{Origin of the gas central to the galaxy}

Having argued that the gas in the filaments originates in gas uplifted from the central galaxy the question then becomes, did the cool and cold gas in the galaxy originate from the cooling hot ICM or through stellar mass loss of the BCG or its principle merger constituents? This is a complicated question and the answer is probably both but whether a dominant mechanism is at play is not known. 

Abell 3581 has a total H$\alpha$ flux of 6.9$\times10^{-14}$ erg cm$^{-2}$ s$^{-1}$; this translates to a mass, M$_{\mathrm{H}^{+}}=2.6\times10^{6}\mathrm{M_{\odot}}$ assuming n$_{e}\sim100$~\pcmcu\ and that the effective recombination coefficient of H$\alpha$ is $\alpha^{eff}=1.17\times10^{-13}$~\cmcups\ \citep{osterbrock2006}. CO$(2-1)$ emission has also been detected in the central galaxy. 
The CO luminosity is 2.35$\times$10$^{7}$K~\kmps~pc$^{2}$; assuming the standard conversion factor (M$_{gas}$/L$^{'}_{CO}=4.6$\Msun(K~\kmps~pc$^{2}$)$^{-1}$) the mass of molecular gas estimated from the IRAM observations of the nucleus of Abell 3581 is $\sim1\times10^{8}$~\Msun. The observable dust mass, estimated from our $E(B-V)$ map, is 10$^{6}$~\Msun.

\cite{sanders2010} find a mass deposition rate of about 0.4 solar masses a year in the central regions of Abell 3581 and a cooling time for this gas of $\sim10^{8}$~\yr. The coolest X-ray gas is morphologically similar to the optical filaments. If these filaments are seeded by being dredged up out of the galaxy by the bubbles then the outer filaments are approximately the age of the outer bubble, $\sim5\times$10$^{7}$ years. In this time, making the extreme assumptions that 100 per cent of the gas cools and that the mass deposition rate has not changed, the X-ray contribution to the filaments could be up to 2$\times10^{7}$~\Msun, easily accounting for the mass of cool gas in the filaments but nearly an order of magnitude shy of the expected cold gas content of the central galaxy. If the filaments are currently growing in mass, some dusty gas must have existed to seed the filaments in the first place. It should be noted that we don't yet know the exact morphology of the cold molecular gas in Abell 3581; however, cold molecular gas has been found coincident with extended ionised filaments in many systems (e.g. \citealt{hatch2005, johnstone2007, oonk2010, mittal2011, salome2011, donahue2011, werner2012}; Werner et al. 2013 in prep.). 

The SFRs given by WISE infra-red and by {\sc galex} near- and far-ultraviolet emission are $\sim$ 0.21~\Msunpyr, 0.3~\Msunpyr\ and 0.1~\Msunpyr\ respectively, close to the current mass deposition rate measured from the cool X-ray gas. However, the morphologies of these two components are very different. The star formation is centred on the nucleus and a single bright young star cluster, and is roughly symmetrical, while the cool X-ray gas is asymmetrically distributed and much of it is associated with the filaments. 

\begin{figure}
\centering
\includegraphics[width=0.45\textwidth]{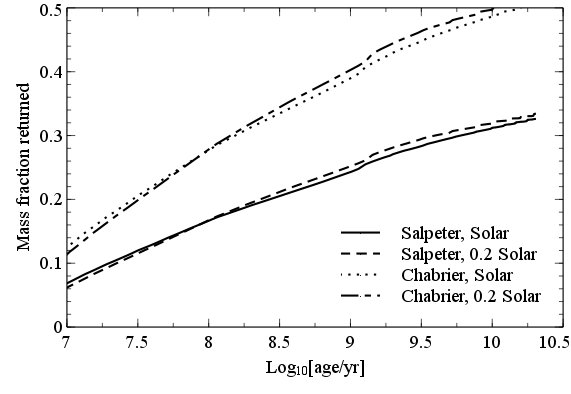}
\includegraphics[width=0.45\textwidth]{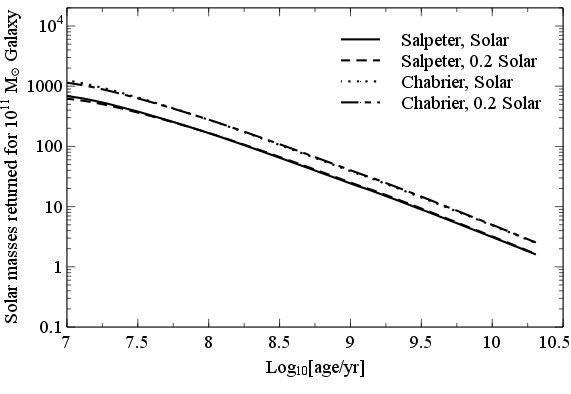}
\caption{The time evolution of mass loss for a simple stellar population (SSP) which has been born at time $t=0$ using the models of BC03 \citep{bc03}. The fraction of the stellar mass returned is very initial mass function (IMF) dependent but only midly dependent on metallicity. \label{ssp_gas_frac}}
\end{figure}

Fig. \ref{ssp_gas_frac} shows that in the lifetime of a BGG a significant fraction of its stellar mass will be returned to the ISM and its various gas phases. The evolved stellar population is currently contributing $\sim$1~\Msunpyr\ (estimate for a 10$^{11}$~\Msun\ galaxy). This is comparable to the cooling rate of the hot gas and the SFR, and as such the stellar mass loss may be an important contributor to the current cool and cold gas being generated in the galaxy. 

We are unable to conclude whether one mechanism dominates over the other; both stellar mass loss and the cooling gas may be providing similar cool gas quantities currently. The stellar mass loss in total, over the lifetime of the population, must be significant, however, how much of the mass is maintained in the cool/cold gas phases is not known. How X-ray mass deposition rates and the heating of the X-ray gas has varied with time is also not known. The more abundant cool and cold gas in BCGs within galaxy clusters with cool cores, as opposed to non-cool core galaxy clusters, suggests that the high pressure environment must play a role. 

\subsection{Filament excitation}

Clues about the filament excitation mechanisms can be gleaned through Fig. \ref{em_ratios}. Near the nucleus there is both a gradient in the [N {\sc ii}]$\lambda$6583 / H$\alpha$ emission line ratio, decreasing rapidly from the nucleus (dropping by a factor of 2 within 5 arcsec), and a gradient from the inside to the outside of the nearby filaments (dropping by a factor of $\sim$1.5 across the filament width). Farther out in the filaments the ratio shows less variation.

At a fixed metallicity and density the [N {\sc
ii}]$\lambda$6583 / H$\alpha$ ratio is essentially an indicator of the amount of
heating in the gas per hydrogen ionization. However, collisional excitation may
be more efficient at higher densities, and increases in the nitrogen abundance in
the gas will increase the [N {\sc ii}]$\lambda$6583 / H$\alpha$ ratio. The
observed gradient could be indicative of an extra
heating mechanism, such as photo-ionisation by the AGN or a weak shock or interaction driven by the AGN
outflow, or a metallicity or density gradient in the gas.

As we suspect the gas originates in the galaxy and has been dragged out and pushed aside by the inflating radio bubbles, strong metallicity gradients over small regions of the filament are unlikely. [O {\sc iii}]$\lambda$5007 is typically weak in the extended filaments in BCGs. Indeed it is too weak to observe in our outer filaments, but is strong in the filament sections close to the nucleus and where the velocity widths of the lines are large. We observe a strong gradient ($>$ factor 3 increase) in [O {\sc iii}]$\lambda$5007 emission from the `inside' of the southern filament, adjacent to the radio bubble, to the `outside' and an accompanying strong gradient in the ratio of [O {\sc iii}]$\lambda$5007 to H$\alpha$ emission. Unfortunately the wavelength range of our spectrum does not include [O {\sc iii}]$\lambda$4363 emission so we are unable to estimate the kinetic temperature variations across the filament.

[S {\sc ii}]$\lambda$6716 / [S {\sc ii}]$\lambda$6731 is a diagnostic of the
electron density. Ratios $>$~1.3 are typical of the low density limit ($<$100~\pcmcu) while the
high density limit is characterised by ratios of $<$~0.5 ($>$1000~\pcmcu). Typically, extended
filaments have been found to lie in the low density limit (e.g.
\citealt{hatch2006}) and this is consistent with our observed values.

We observe emission lines of [N {\sc i}]$\lambda$5199 in the filaments. This emission is usually weak in gas with a well defined kinetic temperature. This is due to the excitation potential being similar in magnitude to the ionisation potential of H$^{0}$. There is often little N$^{0}$ present in warm gas and so little [N {\sc i}]$\lambda$5199 emission (see \citealt{ferland2009, ferland2012} for further discussion of [N {\sc i}]$\lambda$5199 excitation mechanisms). A mechanism which produces supra-thermal particles in the gas and can create some level of ionisation even within cold gas, gas which is multi-phase on a microscopic level, would be successful at reproducing these line ratios \citep{ferland2009}. Such a mechanism, provided by the surrounding hot ISM/ICM, has been suggested by \cite{ferland2008, ferland2009, fabian2011}. Additionally, \cite{churazov2013} have suggested non-thermal particles created in situ in the magnetically supported filaments may contribute to the emission line spectra.

The variation in the emission line ratios shown in Table \ref{emission_lines} highlights the complicated nature of the filament excitation mechanisms in regions which are not isolated, such as those close to the nucleus or close to inflating radio bubbles. Clearly several excitation mechanisms are at play in the inner regions of this filament system and disentangling them is not trivial. It may not be informative to study the emission line ratios in large regions of the centres of these systems; choosing specific `clean' regions where one mechanism dominates is preferable.

\subsection{Jet precession/bulk motions of ICM}

The current bubble axis in IC 4347 has east-west point symmetry through the nucleus. However, the outer bubble appears twisted $\sim$20 degrees south from the westerly direction (in projection); a twist which is also reflected in the trailing H$\alpha$ filaments and the high-to-low frequency radio emission. Lower frequency radio emission probes less energetic plasmas, therefore older bubbles. The effect of bubble wobbling is unlikely to account for the change in the orientation of the radio emission axis as this need not affect both the bubbles in the same way.

Bubble rise directions do not appear to be consistent; some systems exhibit directional changes between outbursts but, for the most part, maintain point symmetry (e.g. NGC1275, \citealt{fabian2012}). \cite{dunn2006} suggested a precessing jet may be responsible for the directional change in the three most recent outbursts of NGC 1275. A motivating feature for this model is that the filaments are largely radial. The bubble direction may also have changed due to turbulence or gas motions within the nuclear region and subsequently risen in the `easiest' direction - that of the lowest pressure - however, it is not obvious that this would lead to bubbles with approximate point symmetry. If the precession model is an explanation for the Abell 3581 bubbles, assuming the bubbles are rising in the plane of the sky, the measured angle between the inner and outer bubbles in Abell 3581 gives a precession rate of 6 degrees per 10$^{7}$~yrs.

NGC 5044 exhibits bubbles which are rising in various directions in the cluster \citep{david2009, david2011}. The authors suggest that the nearly isotropic distribution of the bubbles is a result of the group `weather' from AGN driven turbulence generated by many small outbursts or from the sloshing of NGC 5044 with respect to the group potential (e.g. see simulations of \citealt{bruggen2005, heinz2006}). Finally, a beautiful example of linearly rising bubbles is shown by \cite{randall2011} in their X-ray images of NGC 5813.

Some clues as to the directional change in Abell 3581 are given by observations of the same twist in the optical and radio emission and the large-scale spiral structure observed in the hot X-ray gas beyond the outer bubble (Fig. \ref{xray_outer}). Fig. \ref{cavity_profile} shows evidence for a jump in temperature, a density dip and approximate pressure equlibrium coincident with the outer edge of the spiral, consistent with a cold front.

Similar structures have been observed in many X-ray images of cluster cool cores (e.g. \citealt{churazov2000, fabian2000, mazzotta2001, markevitch2001, markevitch2003, mazzotta2003,
churazov2003, dupke2003, clarke2004, sanders2005, sanders2009}). In relaxed clusters, the current paradigm is that these cold fronts are caused by gas sloshing within the cluster cool core. The sloshing may be a result of a disturbance caused by the in-fall or fly-by of a sub-cluster or alternatively a disturbance associated with energy injection into the ICM by the AGN (for a review see \citealt{markevitch2007}).

\cite{jones2002}, \cite{baldi2009} and \cite{david2009} have found similar spiral features in NGC 4636 and NGC 5044 respectively with the distinguishing feature that these spirals appear to have two arms. Simulations of cluster sloshing have thus far produced one-armed spirals. In NGC 5044 \cite{david2009} suggests the the group weather has been responsible for perturbing the buoyant bubbles rise paths causing them to drag cooler, metal rich material out of the galaxy in two spirals; one emanating from each jet. The `weather' is perhaps still generated by bulk sloshing motions but the cooler gas is now dredged up from the nucleus rather than cooling from the core. NGC 4636 differs in that its spiral arms are hotter, not colder than the surrounding medium suggesting these are the result of shocks driven into the ICM due to the inflation of the bubbles.

In Abell 3581 the X-ray data appears to favour a single spiral structure. The slow rotation in the radio emission and the directional change along the filaments indicates the rise direction differences are not due to jet precession and that some other process changed the bubbles direction after it had left the nucleus. This is unlikely to be simply the path of least pressure as there is a rotational symmetry to the directional change in the filaments. The projected twist direction is clockwise and is the same direction in which the spiral unravels. The stirring of the ICM from some previous disturbance could be responsible for this.

\subsection{Cooling and heating}

Active AGN heating is ongoing in the core of Abell 3581. Bubbles are being produced and rising into the ICM dragging with them multi-phase filaments. The cool and cold gas and dust phases are presumably interacting with the hot surrounding medium resulting in the cool, low volume filling, X-ray emission.

The X-ray luminosity in the core (at a radius of the inner bubbles) is extremely well matched by the powers of the inner bubbles, L$_{X}\sim$2.1$\times$10$^{42}$~\ergps\ and P$_{cav}\sim2.2\times10^{42}$~ergs~s$^{-1}$ respectively. However, assuming the outer bubble to the west of the nucleus is part of a pair, the outer and inner cavity powers, P$_{cav}\sim3.1\times10^{42}$~ergs~s$^{-1}$, are more than a factor of two lower than the X-ray luminosity in a circular region with a radius of the projected distance of the outer cavity (L$_{X}\sim$7.8$\times$10$^{42}$~\ergps) and significantly smaller than the X-ray luminosity within the `cooling radius' (assumed to be where $t_{cool}<7.7$~Gyr, L$_{X}\sim$1.1$\times$10$^{43}$~\ergps). If the bays observed to the south and east are additional bubbles, the total cavity heating within this region may be significantly larger.

The high metallicity, low entropy spiral of gas eminating from the core is indicative of ongoing stripping of the X-ray cool core. This stripping is distributing low entropy material over a larger radius in the cluster in a manner reminiscent of that observed in the much more massive Ophiuchus Cluster \citep{million2010}. The mass of the cool core inside a radius of 30 arcsec ($\sim$13~kpc) is $\sim$10$^{10}$~\Msun. Taking the density from Fig. \ref{spiral_ent} and a conservative cylindrical volume representative of the brightest part of the spiral, with dimensions 2 arcminutes in length by 10 arcseconds in radius, we estimate the mass of the stripped spiral arm to be $\sim$8$\times10^{8}$~\Msun\ (making the assumption of 100 per cent filling factor) or $\sim$8 per cent of the current cool core. Assuming the sound speed of the gas $\sim$400\kmps, the time taken for gas to travel the 2 arcmins from the core to the end of the cylindrical volume in which we have measured the gas mass is $\sim$1.3$\times$10$^{8}$~\yr. Thus, stripping of this rate could in principle destroy the cool core in 1-2~\Gyr, although ongoing cooling (the cooling time in this region is comaprable, $\sim$10$^{9}$~yr) will also replenish it.

\section{Conclusions}

Abell 3581 has a rich multi-phase phenomenon not common in X-ray groups. There is evidence for multiple `radio-mode' outbursts of activity from the central AGN and it is clear the resulting bubbles are interacting with the cold, cool and hot gas in the system. We summarise our main conclusions below:\\

 \noindent $\bullet$ The BGG, IC4373, exhibits bright cool ionised gas filaments extending over 30 arcsec ($\sim$13~kpc) from the nucleus. These filaments are coincident with dust and soft X-ray emission. The filaments engulf the inner east-west bubbles and then change direction extending towards older cavities in the south-west and north-east.\\
 $\bullet$ Based on the morphology of the X-ray cavities, radio emission and cool gas filaments and on the filament kinematics we suggest gas uplifted, by the radio bubbles, from the centre of the galaxy is responsible for the extended, filamentary nature of the ionised gas nebulosity.\\
 $\bullet$ The conclusion, that the filamentary gas originated from the uplift of material from the core of the galaxy rather than the filaments condensing directly from the hot ICM, is supported by the presence of the dust in the filaments which is likely seeded by stellar mass loss in the centre of the galaxy.\\
 $\bullet$ The line emission close to the nucleus and coincident with the direction of the jet has very high velocities indicative of an outflow.\\
 $\bullet$ The BCG has a SFR of about 0.2 - 0.3 \Msunpyr, as measured by WISE and {\sc galex}. A young blue star cluster is also revealed in the HST data. The SFR is comparable to the X-ray mass deposition rate (0.4~\Msunpyr) from the XMM RGS spectral analysis \citep{sanders2010}.\\
 $\bullet$ Farther out on the filament system, the emission line ratios are similar to those seen in the extended filaments of other BCGs exhibiting characteristically low [O {\sc iii}]$\lambda$5007 emission and relatively strong [N {\sc i}]$\lambda$5199 emission. It is probable that many excitation mechanisms are occurring in the `messy' inner regions of the galaxy while the outer `cleaner' regions are consistent with an excitation mechanism which produces some level of ionisation even in cold gas, such as collisions with supra-thermal particles from the hot ISM/ICM. \\
 $\bullet$ The presence of the cold front and the directional change of the bubbles and filaments indicates that the group weather, driven by sloshing, may be controlling the rise direction of the bubbles.\\
 $\bullet$ The spiral cold front exhibits higher metallicities and lower entropy than the surrounding gas and is similar in its X-ray properties to the cool core. The spiral could be material, stripped from the core due to sloshing motions. \\

\section{Acknowledgments}

Support for this work was provided by NASA through award number 1428053 issued by
JPL/Caltech. REAC acknowledges a scholarship from the Cambridge
Philosophical Society and a Royal Astronomical Society grant. 
MS is supported by the NASA grants GO1-12103A, NNH12CG03C and HST-GO-12373.
TEC was supported in part for this work by the National Aeronautics
and Space Administration, through Chandra Award Number
GO1-12159Z. Basic research in radio astronomy at the Naval Research
Laboratory is supported by 6.1 Base funding.
SG acknowledges the support of NASA through Einstein Postdoctoral
Fellowship PF0-110071 awarded by the Chandra X-ray Center (CXC), which
is operated by SAO.
SWA acknowledges support from the U.S. Department of Energy under
contract number DE-AC02-76SF00515.
PEJN was supported by NASA contract NAS8-03060.
CLS was supported in part by NASA Chandra Grant G01-12169X.
We thank the reviewer of this article, Dr G.Tremblay, for his constructive comments and suggestions. REAC would also like to thank Anja von der Linden and Adam Mantz for 
interesting and enlightening discussions.
This research has made use of the NASA/IPAC Extragalactic Database (NED) which
is operated by the Jet Propulsion Laboratory, California Institute of
Technology, under contract with the National Aeronautics and Space
Administration.

\bibliographystyle{mnras}
\bibliography{\dirlatexcommon mnras_template}

\end{document}